\documentclass[inte,nonblindrev]{informs3} 

\OneAndAHalfSpacedXII 

\usepackage{natbib}
 \bibpunct[, ]{(}{)}{,}{a}{}{,}%
 \def\bibfont{\small}%

\usepackage{multibib}
\usepackage{chronosys}
\newcites{AP}{References for the Appendix}
\newcommand{\tabitem}{~~\llap{\textbullet}~~}
\TheoremsNumberedThrough     

\EquationsNumberedThrough    
\usepackage{amsmath, bbm}
\usepackage{float}
\usepackage{multirow}
\usepackage{xcolor}
\MANUSCRIPTNO{OM-10-2023-0076} 

\usepackage[suppress]{color-edits}
\addauthor{yz}{magenta}
\addauthor{bl}{blue}
\addauthor{sh}{red}
\addauthor{edit}{red}
\newcommand*\samethanks[1][\value{footnote}]{\footnotemark[#1]}
\begin{document}


\RUNAUTHOR{Liu et al.}

\RUNTITLE{Modeling the Risk of In-Person Instruction during the COVID-19 Pandemic}

\TITLE{Modeling the Risk of In-Person Instruction during the COVID-19 Pandemic}

\ARTICLEAUTHORS{%
\AUTHOR{Brian Liu\thanks{Equal contribution.}}
\AFF{Operations Research Center, Massachusetts Institute of Technology, \EMAIL{briliu@mit.edu}}
\AUTHOR{Yujia Zhang\samethanks}
\AFF{Center for Applied Mathematics, Cornell University, \EMAIL{yz685@cornell.edu}}
\AUTHOR{Shane G. Henderson}
\AFF{School of Operations Research \& Information Engineering, Cornell University, \EMAIL{sgh9@cornell.edu}}
\AUTHOR{David B. Shmoys}
\AFF{School of Operations Research \& Information Engineering, Cornell University, \EMAIL{dbs10@cornell.edu}}
\AUTHOR{Peter I. Frazier}
\AFF{School of Operations Research \& Information Engineering, Cornell University, \EMAIL{pf98@cornell.edu}}
} 

\ABSTRACT{%

During the COVID-19 pandemic, safely implementing in-person indoor instruction was a high priority for universities nationwide. To support this effort at the University, we developed a mathematical model for estimating the risk of SARS-CoV-2 transmission in university classrooms. This model was used to evaluate combinations of feasible interventions for classrooms at the University during the COVID-19 pandemic and identify the best set of interventions that would allow higher occupancy levels, matching the pre-pandemic numbers of in-person courses, despite a limited number of large classrooms. 
\editedit{Importantly, we determined that requiring masking in dense classrooms with unrestricted seating when more than 90\% of students were vaccinated was easy to implement, incurred little logistical or financial cost, and allowed classes to be held at full capacity.}
A retrospective analysis at the end of the semester confirmed the model's assessment that the proposed classroom configuration would be safe. 
Our framework is generalizable and was used to support reopening decisions at Stanford University. 
In addition, our framework is flexible and applies to a wide range of indoor settings. It was repurposed for large university events and gatherings and could be used to support planning indoor space use to avoid transmission of infectious diseases across various industries, from secondary schools to movie theaters and restaurants.

}



\maketitle

%



\section{Introduction}\label{intro}

During the initial period of the COVID-19 pandemic, from March 2020 to May 2021, many universities switched entirely to virtual instruction because of a fear that a large outbreak in the student population could quickly overwhelm local healthcare capacity and endanger students, employees, and residents who live near campus \citep{walke2020preventing, cipriano2021impact}. 
These interventions were not without costs, as they harmed the social well-being and educational outcomes of college students \citep{lee2021impact, dorn2020covid} and damaged the local economies of college towns \citep{collegetown_financial_impact, sullivan2020college}. 
Moreover, prolonged campus shutdowns negatively impact student learning \citep{dorn2020covid} and the livelihoods of those who work around campus \citep{sullivan2020college}. Therefore, safely reopening college campuses to accommodate in-person instruction while avoiding the transmission of infectious diseases is important for universities nationwide.


The University in Ithaca, New York (name removed to support blinding) was a leader in safely re-opening for residential instruction \citep{frazier2022modeling}. In the Fall of 2020, over 75\% of all students enrolled at the Ithaca campus returned for in-person instruction \citep{dailysun_fall2020} and extensive testing, contact tracing, and classroom de-densification protocols resulted in fewer than 200 COVID-19 cases throughout the semester out of a population of over 18,000 students \citep{feb2021_report}.
During this semester, however, while two-thirds of students had at least one in-person class \citep{dailysun_fall2020}, only 30\% of all courses were held in-person \citep{dailysun_fall2020_classes_stats}. 
A mandated six-feet distancing requirement, set by the New York State Department of Health \citep{nys_distancing}, constrained the number of students that could be accommodated in each classroom. For example, a class with 200 students required a classroom that seated 1600 people. This mandate dramatically reduced the number of rooms on campus that could accommodate a large class of students. Furthermore, rooms with poor air circulation were excluded from usage, and only a limited number of classrooms could be retrofitted with HVAC to enhance ventilation due to high operational and energy costs. As a result, it was impossible to schedule many classes in person.



While the Fall 2020 semester proved that the University had the ability to safely reopen campus, and the level of in-person instruction during that semester was substantially above that offered by many other universities at the time \citep{fall_2020_ivies_plans}, the number of in-person classes remained significantly below pre-pandemic levels. This continued in Spring 2021; the number of in-person courses offered remained lower than pre-pandemic levels and again most students who returned to campus enrolled in hybrid schedules and took the majority of their classes virtually \citep{dailysun_spring2021_classes}.



\begin{figure}[h]
    \centering
    \caption{Timeline of significant events during the planning period for the Fall 2021 Semester. }
    \includegraphics[width = \textwidth]{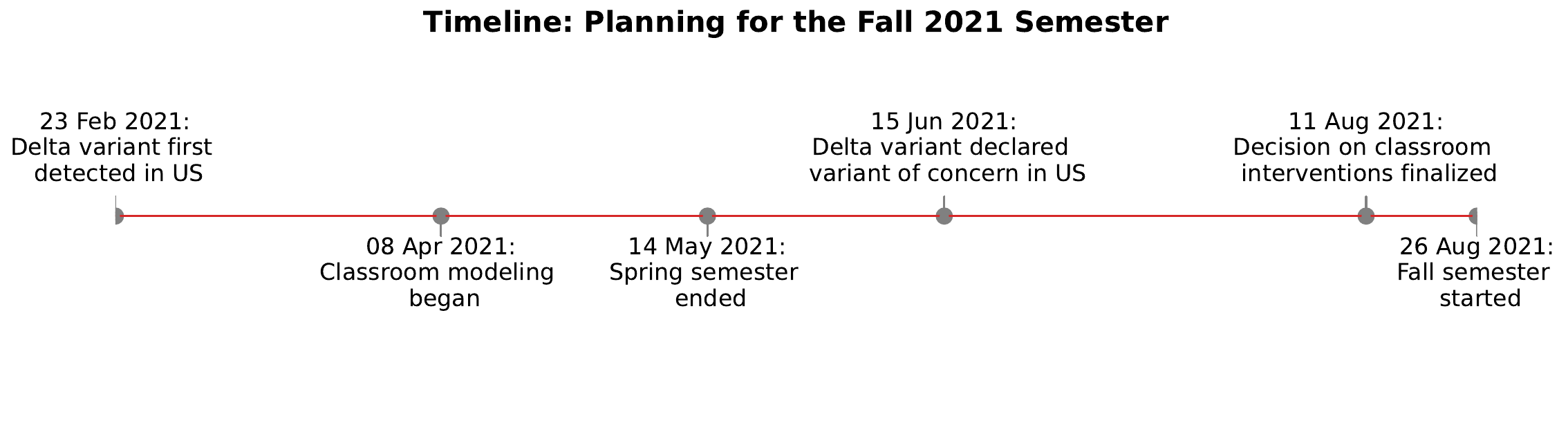}
    \label{FA2021.timeline}
    {\small \textit{Note}: The Delta variant was declared a variant of concern in the United States halfway through the planning process and influenced our modeling approach and decisions.}
\end{figure}

The University started to gauge the possibility of offering the full roster of courses in-person when planning the Fall 2021 semester, since much of the community would have been vaccinated at the onset of the semester. However, the University faced considerable uncertainty about the safety of offering a full roster of in-person courses. The level of safety associated with using all classrooms, not just rooms with high-quality ventilation, and filling them at greater density was not well understood. Adding to this challenge, the SARS-CoV-2 Delta variant emerged in the summer of 2021 with increased infectivity compared to the original strain and resistance to vaccines \citep{callaway2021delta}. Figure \ref{FA2021.timeline} shows a timeline of how the emergence of the Delta variant coincided with our planning period for the Fall 2021 semester.


To respond to this uncertainty, our team developed a modeling framework to estimate the risk of COVID-19 transmission in classrooms during the Delta wave of the pandemic. Using this framework, we determined that fully dense classrooms with mandatory masking and without special ventilation or restrictive seating plans would result in minimal risk to students, graduate student instructors, faculty, and teaching staff throughout the semester. Thus, safe in-person instruction could be offered without further enhancing ventilation in classrooms or developing fixed seating plans for each class,
interventions that would have been difficult to implement.
Following our recommendations, the University proceeded with dense in-person classes in Fall 2021, and empirical evidence aggregated at the end of the semester suggested that classroom transmission was extremely rare \citep{dec2021_end_of_semester_guidance}. 

Our modeling framework can be used to support the design of interventions during respiratory disease outbreaks in any context with indoor seating, from K-12 schools to restaurants and movie theaters. Our framework is flexible and allows a user to estimate the risk of virus transmission in rooms with various configurations. In addition, our framework can evaluate the effectiveness of various interventions, such as vaccines, masking, and ventilation. These functionalities make our framework a valuable tool for modeling indoor transmission. 


\subsection{Contributions to the University}\label{contribution}



Our modeling framework and analysis guided decision makers at the University in planning for the difficult task of resuming normal teaching operations for the Fall 2021 semester. We used our framework to recommend an implementable classroom configuration that allowed classes to meet in full density while ensuring safety for students and instructors.
The classroom policies that we recommended, namely mandatory masking with no distancing or additional ventilation requirements, sufficiently prevented COVID-19 transmission in classrooms. Our retrospective analyses (including contact tracing of COVID-positive students and employees, adaptive testing of students in classrooms with positive cases, and genetic sequencing of viral samples) at the end of the semester found that student travel and social events were much more influential drivers of COVID-19 spread on campus compared to classroom transmission \citep{track_transmission_webarchive}.

 We also communicated our modeling approach with transparency and rigor through published analyses and town hall meetings \citep{fall2021_classroom_modeling_report,faculty_townhall}. Prior to the start of the Fall 2021 semester, many in the community expressed concern about the safety of in-person classes. In August 2021, the University Chapter of the American Association of University Professors expressed in a letter to the University President concerns about the risk of teaching in person due to the increased transmissibility  of the Delta variant \citep{AAUP_letter}. During a faculty and staff town hall the same month, multiple questions were asked about the risk of transmission from teaching class and holding office hours, masking in classrooms, and the efficacy of vaccines against the Delta variant \citep{faculty_townhall}. In addition, multiple faculty members emphasized that the University needed to be more transparent about how classroom safety was assessed \citep{faculty_letters}.
 
 We developed and communicated our modeling framework to reassure the community that in-person instruction was safe using transparent, data-driven methods.
As a result, the university was able to more effectively communicate and inform instructors, teaching assistants, students, and the broader community that returning to normal teaching operations had minimal risk.


Beyond classrooms, we also used our modeling framework to evaluate the risk of holding and attending other university events, such as Homecoming, concerts, holiday events, sporting events, and graduation. These analyses informed executive-level decisions on which events to hold throughout the Fall 2021 semester.
As a result, we found our modeling approach to be useful for evaluating the risk of virus transmission in many indoor settings.

Our modeling framework and analyses were widely distributed and influenced return-to-campus decisions at other universities. Notably, Stanford University cited our analyses in their decision to return to on-campus instruction for Fall 2021 \citep{stanford_2021_cite_cornell}. We believe that our success, along with the flexibility and generalizability of our modeling approach, makes our framework a useful tool for managing indoor operations during respiratory disease outbreaks and a valuable contribution towards mitigating the impacts of pandemics.

\subsection{Related Work}


\editedit{
Mathematical modeling was crucial for supporting college reopening decisions during the pandemic.
In 2020, universities employed optimization tools in designing course schedules to satisfy multiple decision criteria \citep{navabi2022multicriteria} such as minimizing student interactions \citep{gore2022clemson} and maximizing the number of in-person courses offered \citep{johnson2022practice}.
These modeling approaches allowed universities to resume in-person instruction in limited capacities in accordance with social distancing regulations. 

Multiple studies have evaluated the risk associated with classroom instruction. 
These models either use high-fidelity, yet time-consuming, computational fluid dynamics simulation \citep{foster2021estimating,mohamadi2022review}, or simulate airborne transmission probabilistically without accounting for the spatial locations of susceptible students \citep{hekmati2022simulating,ucboulder_spreadsheet,bazant2021guideline}.
Our work provides value in that we assessed the risk of in-person instruction when social distancing regulations were relaxed, an important consideration when returning to pre-pandemic levels of in-person instruction.
We model the spatial variation in transmission in a tractable way. Coupled with quantification of parameter uncertainty, our framework provides efficient and robust assessment of different classroom settings in practical situations. 
}


The rest of this paper is organized as follows. First, we describe in detail the challenges faced by the  University when planning for the Fall 2021 semester. We then explain our framework for estimating the risk of COVID-19 transmission in classrooms, which includes mathematical modeling and a computer simulation. We apply our framework to evaluate different interventions and develop a strategy to safely operate dense in-person classrooms that was recommended to university leadership. We conclude with a retrospective evaluation of our model's validity and discuss its broader impact beyond modeling transmission in classrooms. Further details of our model are presented in the appendix. 

\section{Problem Statement} \label{problemstatement}







While planning for the Fall 2021 semester, the University aimed to offer as many in-person classes as possible while maintaining classroom safety. 
For the Fall 2020 and Spring 2021 semesters, the University had held only a limited number of de-densified classes, where the students were spaced six feet apart. 
The constraint of having a finite number of classrooms on campus posed a challenge, as expanding in-person classes would elevate student density in classrooms, potentially heightening the risk of indoor COVID-19 transmission.
To mitigate this potential for elevated transmission risk, the University needed to implement classroom interventions. Interventions under consideration included requiring masking, improving ventilation, increasing social distancing, and assigning seats randomly. (Assigning seats randomly would reduce the risk that unvaccinated students, who are more vulnerable to infection and have higher transmission when infected, would sit together in socially connected groups.) At the time, we had a limited understanding of the effectiveness of these interventions in preventing disease spread, whether deployed individually or combined. 
Amidst such uncertainty, one major goal of our modeling work was to identify a combination of interventions to efficiently curb disease transmission within classrooms, all while maintaining a reasonable cost.

The University also faced additional concerns in the months leading up to the Fall 2021 semester. The more infectious Delta variant of COVID-19 was spreading globally and was responsible for a deadly second wave in India \citep{tareq2021impact}. There was concern that the variant would spread to the United States and quickly become the dominant strain. Though many students were fully vaccinated, the vaccine's efficacy against Delta was uncertain.
Therefore, 
the goal of our modeling work was to understand what classroom interventions were needed to safely hold dense in-person classes and to assess and communicate how these interventions addressed the concerns that we faced heading into the Fall 2021 semester. We further discuss classroom density, classroom interventions and the Delta variant below.

\subsection{Classroom Density}

\begin{figure}[h!]
    \centering
    \caption{Floor plan of Olin 155, a large lecture hall at the University. }
    \includegraphics[width=0.85\textwidth]{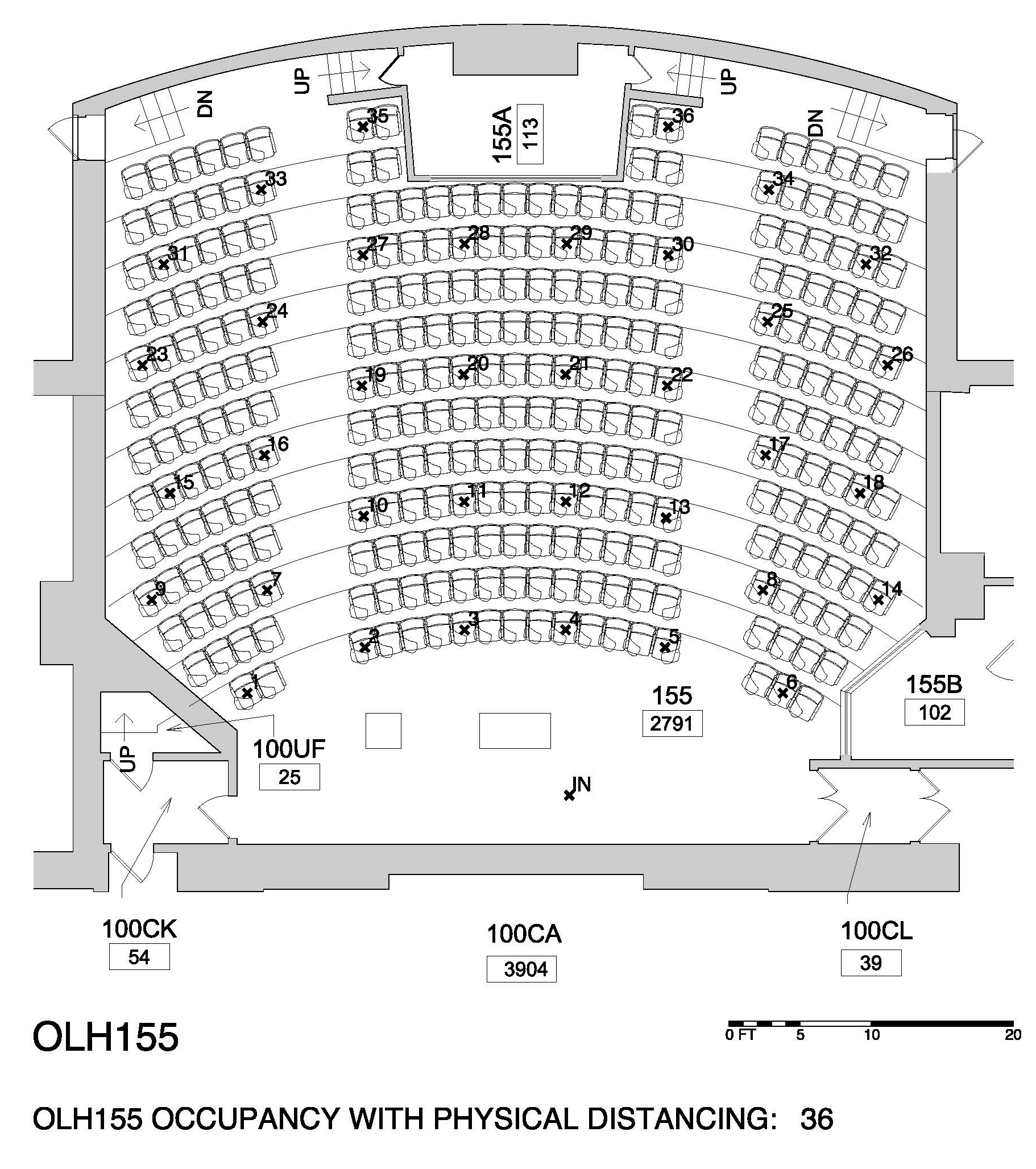}\\
    \label{floorplan.fig}
    {\small\textit{Note}: The socially distanced seating configuration used in Fall 2020 is marked with ``x"; only 36 seats were used among the 287 seats available.}

\end{figure}

During the Fall 2020 semester, only 20\% of courses were offered in-person. While the majority of students returned to Ithaca, few students were in classrooms on any given day during the semester. As such, the University had the ability to aggressively de-densify classrooms to reduce the potential for in-person transmission. All classrooms were configured to be socially distanced, where students were seated six feet apart \citep{CTRO_seating}. Figure \ref{floorplan.fig} shows the floor plan of a socially distanced classroom, Olin Hall 155, that could normally accommodate 287 students during normal university operations. In Fall 2020, with social distancing, the hall could only accommodate a maximum class size of 36 students, a 90\% reduction from the pre-pandemic capacity.

Overall, social distancing reduced campus-wide classroom capacity by 87\%. This reduced capacity was sufficient for the Fall 2020 semester, where only a fraction of courses were offered in-person under reduced schedules.
However, maintaining the same distancing level for the increased in-person course schedule for Fall 2021 would have required each room to be used for more than 24 hours each day.
Thus, further analysis was necessary to assess the safety of increasing classroom density.



\subsection{Classroom Interventions}


The University considered a set of potential interventions to improve classroom safety that included requiring masking, improving ventilation, increasing social distancing, and assigning seats randomly in classrooms. These interventions faced varying implementation difficulties (Table~\ref{tab:interventiondifficulty_table_2}). For example, requiring masking was the easiest intervention to execute since the requirement could be enacted impromptu by the administration. Assigning seats randomly required advance planning to develop seating plans before the start of the semester. These randomized seating plans reduce the chance that unvaccinated students, who have higher susceptibility and transmissibility when infected, sit together in groups. It was even more difficult to increase social distancing since doing so reduces classroom capacity and limits the number of in-person courses offered. Classes would also need to meet at inconvenient times (late night or early mornings) to accommodate  reduced classroom capacity. Finally, increasing ventilation in classrooms was the most difficult intervention to implement, due to the cost of retrofitting all classrooms with HVAC equipment. Such improvements were only made in Summer 2020 for the largest classrooms at the University (classrooms with more than 100 seats that were used for socially-distanced instruction in Fall 2020 and Spring 2021).

In Appendix A2, we describe how we modeled classroom interventions to estimate their efficacy before the start of the semester. This analysis informed the university on interventions needed to ensure safety.

\begin{table}[h]
\centering
\caption{Table of potential interventions and associated implementation difficulty. Medium and hard level interventions must be planned out months before the start of the semester.}
\scalebox{0.7}{
\begin{tabular}{|c|c|c|c|}
\hline
\textbf{Intervention/Difficulty} & Easy                                                                                           & Medium                                                                                              & Hard                                                                                                                      \\ \hline
Require Masking                  & \begin{tabular}[c]{@{}c@{}} \tabitem Instant implementation.\\ \tabitem No effect on class capacity.\end{tabular} &                                                                                                     &                                                                                                                           \\ \hline
Implement Seating Policy         &                                                                                                & \begin{tabular}[c]{@{}c@{}} \tabitem Time consuming to implement.\\ \tabitem No effect on class capacity.\end{tabular} &                                                                                                                           \\ \hline
Increase Distancing            &                                                                                                & \begin{tabular}[c]{@{}c@{}}\tabitem Time consuming to implement.\\  \tabitem Reduces class capacity.\end{tabular}      &                                                                                                                           \\ \hline
Increase Ventilation           &                                                                                                &                                                                                                     & \begin{tabular}[c]{@{}c@{}} \tabitem Time consuming to implement\\ \tabitem Expensive equipment.\\ \tabitem No effect on class capacity.\end{tabular} \\ \hline
\end{tabular}
}
\label{tab:interventiondifficulty_table_2}
\end{table}

\subsection{Delta Variant Uncertainties} 

\begin{figure}[h!]
    \centering
    \caption{Daily COVID-19 cases counts for New York State based on reports from state and local health agencies \citep{times_2021}.} 
    \includegraphics[width=0.85\textwidth]{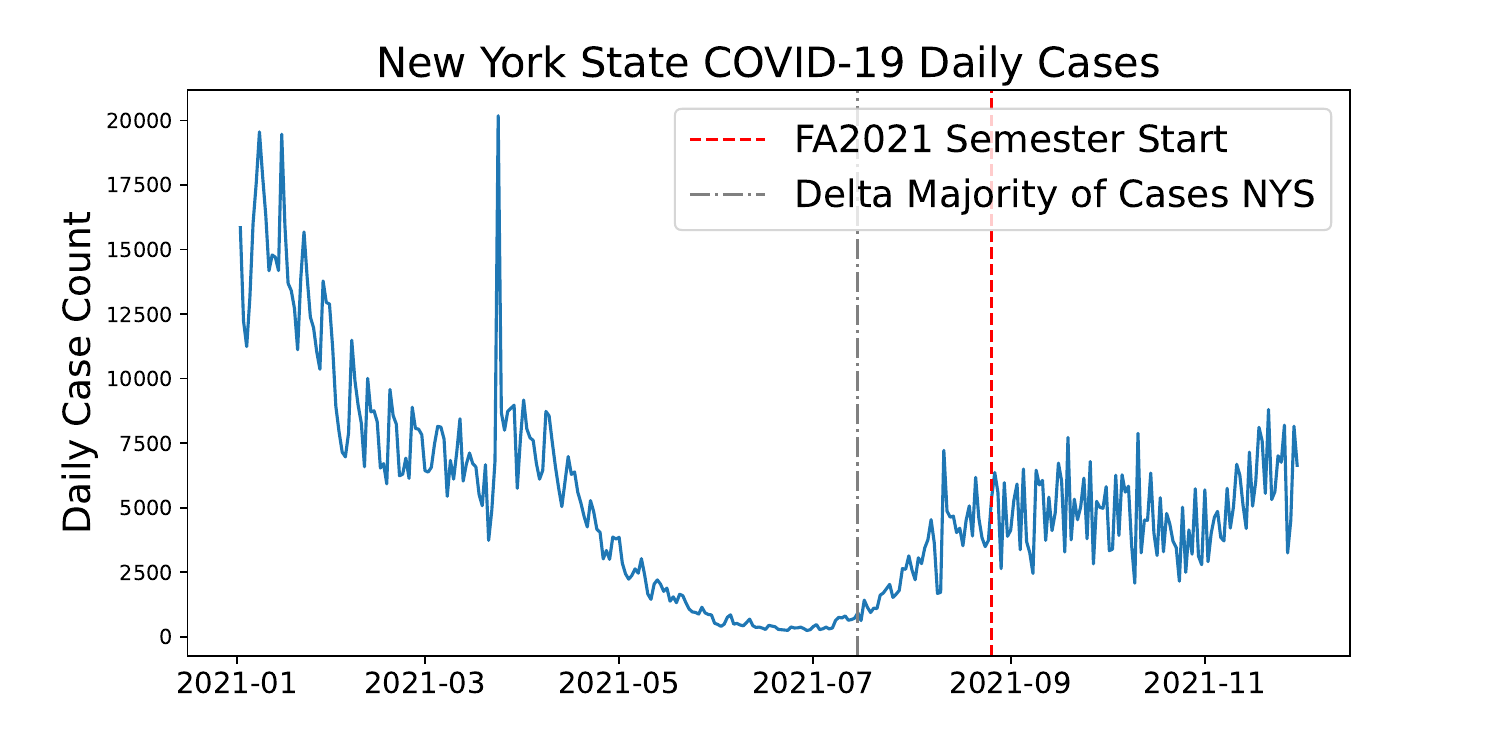}
    \label{NY-covid-cases}\\
    {\small\textit{Notes}: The gray dot-dashed line shows when the majority of cases in New York City were first determined to be from the Delta variant. The University started its Fall 2021 semester roughly 2 months later, at the peak of the Delta wave.}
\end{figure}

Figure \ref{NY-covid-cases} shows daily COVID-19 case counts in New York State in 2021. The dot-dashed line indicates the first date where the majority of cases in New York City were determined to be from the Delta variant  \citep{variantNYS}. The total daily case count in the state rose steadily from that date until the start of the Fall 2021 semester, indicated by the red dashed line in the figure. In retrospect, it is apparent that the semester started during the peak of the Delta wave of the pandemic.

The emerging Delta wave presented challenges when planning for the Fall 2021 semester. First, while the Delta variant drove an increase in cases during the summer of 2021, the exact increase in Delta's infectivity compared to the original strain was not well understood at the time when the University needed to decide on classroom density. 
In addition, the literature on vaccine efficacy against the Delta variant was sparse. Preliminary reports from the United Kingdom and Israel were not encouraging; early studies from the UK NHS \citep{andrews2021effectiveness} and Israel Health Ministry \citep{covid_delta_israel} estimated BNT162b2 Pfizer vaccine efficacy of $88\%$ and $39\%$ respectively against symptomatic illness from the Delta variant. In context, the BNT162b2 Pfizer vaccine achieved vaccine efficacy of $95\%$ against the original strain in clinical trials \citep{polack2020safety}.  


Appendix A2 explains how we estimated the Delta variant's increased infectivity and decreased vaccine efficacy to produce models of classroom risk robust to uncertainty in both parameters. We used these models to determine if the University could safely hold in-person classes during the Delta wave of the pandemic.




\section{Modeling Framework}





The modeling framework we developed consists of two parts: a mathematical model used to estimate transmission risk between individuals under different conditions and a simulation tool used to evaluate overall classroom risk. 
We sketch the main ideas here and provide a full description of our methodologies in the appendix. 

\subsection{Main Assumptions and Parameters}


Our models rely on a set of parameters, the values of which are key to the predictions, and we estimated parameter values from the literature available at the time of our analysis. 
For parameters with high uncertainty, we imposed reasonably chosen prior distributions on their values rather than using point estimates. 
Our assumptions were influenced by our previous work on developing epidemiological models for COVID-19 at the University \citep{frazier2022modeling}. 

We assumed that the Delta variant would be dominant at the University at the start of the Fall 2021 semester, and would be 2.4 times as transmissible as the original COVID-19 strain \citep{washington2021emergence,callaway2021delta}. 
We conservatively estimated that $90\%$ of the undergraduate population would be fully vaccinated at the start of the semester. Among the vaccinated population, we estimated the distribution of vaccine efficacy (VE) against infection and VE against transmission to be centered around 52\% and 51\% respectively. These estimates were obtained by weighting the results from several different studies by their sample size. 
We estimated that if either the source or the susceptible person were masked, the transmission probability would be reduced by 50\% to 80\%.
Finally, we assumed perfect compliance with any masking guidelines given by the University, since, in previous semesters, compliance to COVID-19 regulations was very high.


\subsection{Mathematical Model of Transmission}

Given an infectious person in a classroom, 
we decomposed the risk of transmission into a short-range and a long-range component, each representing a major mode of SARS-CoV-2 transmission \citep{cdc_transmission_modes}. The short-range component models transmission due to the deposition of virus-containing respiratory droplets onto exposed mucous membranes; the long-range component models transmission due to the inhalation of virus-containing aerosols or fine droplets. 
In both components, we used an exponential dose-response model \citep{watanabe2010development}, where the dose is the amount of virus a susceptible individual is exposed to. 
The likelihood a susceptible individual becomes infected, given dose $D$ and a positive constant $c$, can be expressed as,
\begin{equation}
    \mathbb{P}(\text{transmission}) = 1 - \exp(-c\cdot D).\label{eq:expdose}
\end{equation}
As the dose increases, the likelihood of transmission increases as well. 

\subsubsection{Short-range Transmission}

In short-range transmission, the source exhales virus-containing droplets, which are large, heavy particles that tend to deposit on the ground or other surfaces. As the droplets are heavy and cannot travel far, the concentration of droplets in the air decreases with the distance from the source case \citep{mittal2020flow}. 


To model the fact that students mostly face the instructor, who typically stands in the front of the classroom, we assumed the source case emits virus particles in a cone of directions towards the front; we call this set of directions the source case's \textit{cone of exposure}. We modeled the transmission probability in two dimensions, accounting for the distance and angle of the susceptible individual relative to the source case.


We used maximum likelihood estimation to fit the model parameters (including the angle of the cone of exposure) based on a large dataset on COVID-19 transmission aboard high-speed trains in China \citep{hu2021risk}, assuming that all secondary infections in the data were due to short-range transmission.
The dataset gave us the relative positions between infectious index cases on the train and nearby susceptible passengers, as well as the subsequent case incidence rates among the susceptible passengers.
The seating configuration of the train car is similar to a lecture hall, where all individuals face the same direction and are spaced apart by rows of seats.
To the best of our knowledge, this dataset was the best  available at the time we fit our model.

\subsubsection{Long-range Transmission}
We used the model and parameters in \cite{schijven2021quantitative} and modeled long-range transmission by quantifying the concentration of virus-containing aerosols or fine droplets suspended in the air (hereafter, we call them ``aerosols"). 
The model assumed that aerosols are distributed uniformly across the room. 
As a result, the probability of transmission does not depend on distance or angle from the source and only depends on the rate of aerosol emission from the infectious source, duration of exposure, room volume, and level of ventilation. 

\subsubsection{Overall Risk}
We combined the estimated short-range and long-range transmission risk by taking the larger of the two. 

When estimating the parameters for the short-range model, we assumed that all secondary infections in \cite{hu2021risk} were due to short-range transmission, while in reality some cases may have arisen from long-range transmission. Therefore, the estimates for the short-range model may implicitly include some effect of long-range transmission. Setting the overall risk to the maximum, rather than the sum, of the two risks prevents overestimation. In fact, the simulated short-range risk was usually one to two orders of magnitude larger than the long-range risk within three meters, so it dominated the overall risk for those exposed to it. 
This is consistent with \cite{ontairo2022}, which found that shorter distance usually implies higher transmission risk.


We assumed that instructors are sufficiently distanced from the students such that short-range transmission is not possible. In our model, the risk from short-range transmission is negligible after six feet of distancing and we assumed, based on prior semesters, that most instructors spend the majority of their time over six feet away from students.

We did not explicitly model an infectious instructor, because case investigations in the 2020-21 academic year did not reveal any faculty or student infections that were linked to classroom-based transmission \citep{faculty_townhall}, and faculty prevalence was much lower than that of students. In addition, the number of students in a class was typically much larger than the number of instructors. Moreover, even if the instructor was infectious in addition to an infectious student in the classroom, this merely approximately doubles the risk due to long-range transmission for each susceptible student. For the susceptible students most at risk, i.e., those sitting in the proximity of the infectious student, the risk from short-range transmission dominates that from long-range transmission by two orders of magnitude. Therefore, the expected number of secondary transmissions remains almost the same regardless of the instructor's infection status.

\subsubsection{Reflections} 
We developed this modeling framework in Summer 2021 to support reopening decisions at the University for the Fall 2021 semester. As the body of COVID-19 related literature expands, we recommend these modifications to our framework for future use.

\begin{enumerate}
    \item Evaluate and compare other theoretical models for estimating the risk of COVID-19 transmission through droplets and aerosols \citep{mirzaei2021simplified, bazant2021guideline}. In addition, calibrate the model to more datasets that shed light on COVID-19 transmission in enclosed spaces, such as in restaurants \citep{cheng2022outbreak} and theaters \citep{adzic2022post}, as well as adjust for more recent variants such as Omicron \citep{ji2022increasing}.

    \item Update the estimates of virus transmissibility and vaccine efficacy based on the most up-to-date findings \citep{ciotti2022covid, wan2023booster}.
    \end{enumerate}

\subsection{Classroom Simulation Tool}

\begin{figure}[h!]
    \centering
    \caption{Example illustration of classroom simulation tool. } 
    \includegraphics[width=0.85\textwidth]{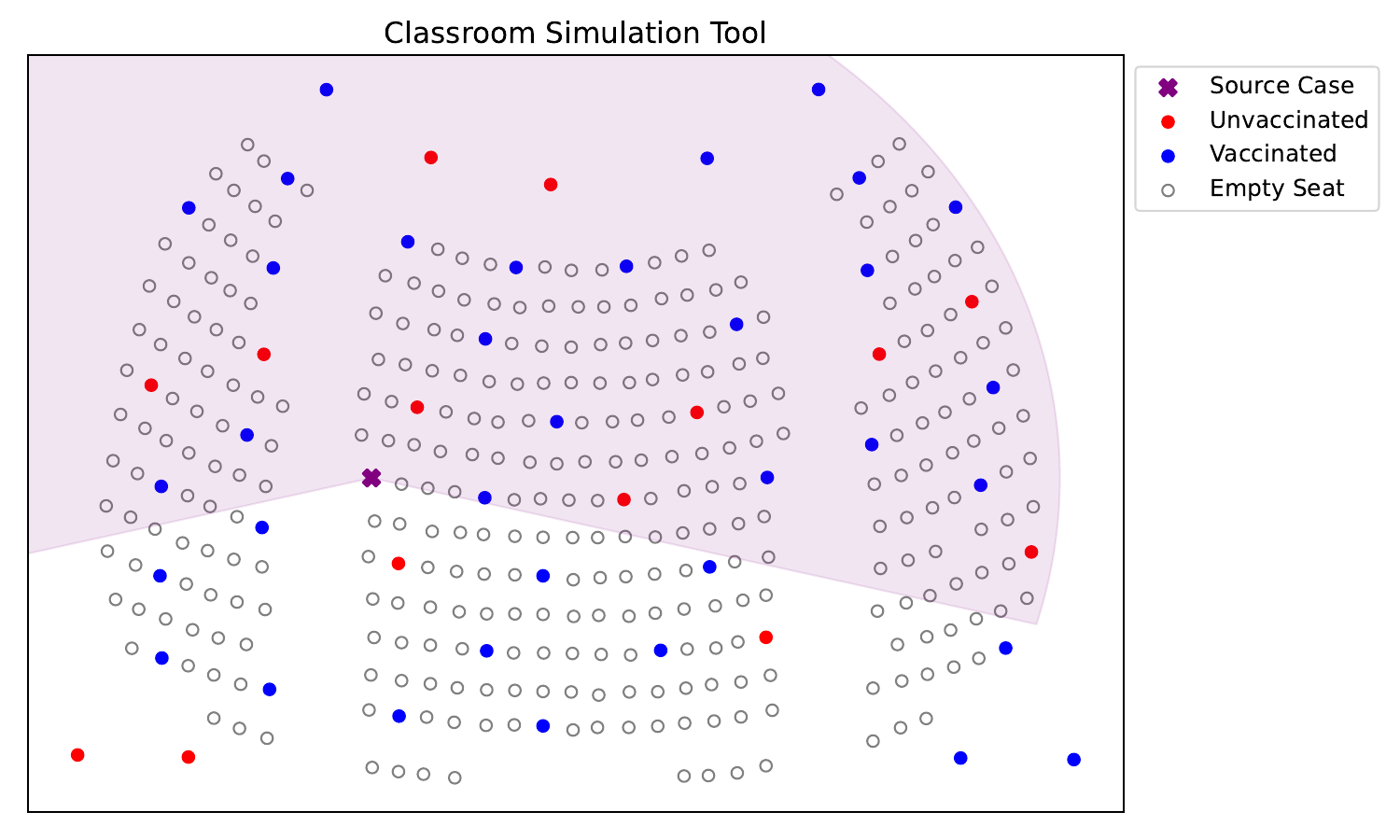}
    \label{classroomsimtool.fig}\\
    {\small\textit{Notes}: The purple \textbf{X} indicates the infectious source case and the purple cone indicates the cone of exposure, the set of directions in which the source case is modeled as emitting virus particles. Unvaccinated and vaccinated students are represented with red and blue dots respectively. The instructor is located on a stage sufficiently distanced from the class, far above the top margin of the illustration.}
\end{figure}

In conjunction with our mathematical classroom model, our simulation tool allowed us to estimate the risk of classroom transmission along with the effectiveness of various interventions, such as masking, social distancing, and increased ventilation. Figure \ref{classroomsimtool.fig} presents an illustration of the classroom simulation tool for a large lecture hall.


For each parameter setting (density level, vaccination rate, vaccine efficacy), we estimated the expected number of secondary infections in the classroom over a 1-hour period given one infectious source case among 50 students, averaged over 500 trials. (We omitted the scenarios where there were two or more source cases in the same classroom at the same time. Such scenarios were unlikely compared to scenarios with one source case because prevalence was low, so they contributed little to overall risk. Further discussion is given in Appendix A1.) For each trial, we randomly generated a seating configuration and vaccination statuses among the students and randomly drew a student to be the source case. We repeated this for all combinations of density level, vaccination rate, and vaccine efficacy. We assumed everyone was unmasked in the simulation. The effect of masking, modeled as an uncertain parameter with a normal prior, can be directly imposed on the results above through multiplication.






\subsection{Interventions and Scenarios Evaluated}
\label{interventionsframework}

\begin{table}[]
\centering
\caption{Effectiveness of intervention methods at reducing short and long-range transmission.}
\begin{tabular}{|c|cc|}
\hline
Intervention/Reduction in transmission & \multicolumn{1}{c|}{Short-range}    & Long-range      \\ \hline
Masking      & \multicolumn{1}{c|}{\checkmark} & \checkmark     \\ \hline
Seating Policy   & \multicolumn{1}{c|}\checkmark & \multicolumn{1}{c|}{-} \\ \hline
Distancing   & \multicolumn{1}{c|}\checkmark & \multicolumn{1}{c|}{-} \\ \hline
Ventilation  & \multicolumn{1}{c|}{-}                    & \checkmark \\ \hline
\end{tabular}
\label{tab:interventions_for_reducing_transmission}
\end{table}

Combining the mathematical model and classroom simulation tool, we evaluated several interventions (masking, seating policy, distancing, ventilation) across different scenarios. Table~\ref{tab:interventions_for_reducing_transmission} summarizes the possible interventions along with their effectiveness against short-range and long-range transmission. 
We discuss these interventions and scenarios in further detail below.

\subsubsection{Masking:}
Based on experimental and observational studies, we assumed the masking effectiveness against transmission to range from 50\% to 80\% if either the infectious or the susceptible individual were masked (see details in Appendix A2). If both of them were masked, the risk of transmission is reduced by 75\% to 96\%. 

We evaluated the intervention of masking for the entirety of the Fall 2021 semester, and we assumed that there was perfect compliance with the masking mandate, consistent with the high compliance observed in previous semesters \citep{faculty_townhall}. 

\subsubsection{Seating policy:}
We considered two different seating policies: (1) randomly assign students to seats and enforce that students always sit in their assigned seats (\textit{fixed seating}); (2) allow students to sit wherever they want (\textit{unrestricted seating}). 

Unrestricted seating had the potential to be more risky in that unvaccinated students could potentially group together. This would lead to a higher expected number of transmissions since unvaccinated students were more susceptible to COVID-19 infection and had a higher transmissibility if infected \citep{de2021vaccine,lopez2021effectiveness}. 


On the other hand, fixed seating was operationally difficult to implement. Our initial simulations showed that the fixed and unrestricted seating policies lead to comparable risk (see Figure~\ref{fig:seating_policies_comparison} in Appendix A1). As a result, the University adopted the unrestricted seating policy; all simulation results shown here are thus based on unrestricted seating.






\subsubsection{Social Distancing:}
We evaluated three social distancing options. In \textit{fully dense} seating, the default spacing for lecture halls pre-pandemic, students were distanced one foot apart from each other in the classroom. In \textit{moderately dense} seating, students were distanced three feet apart. In \textit{distanced} seating, students were seated six feet apart. This last configuration was used in the 2020-2021 academic year during the pandemic.

\subsubsection{Ventilation:}

For an infectious source case, we assumed that aerosol viral particles were emitted continuously over the hour at a constant rate and were immediately distributed evenly across the room once emitted. We quantified ventilation rate by measuring how often air was exchanged from the room, in the unit of air exchanges per hour (ACH), and we assumed that air exchanges happen evenly over time. According to the University's Facilities Department, most classrooms had a ventilation rate of 1 ACH. We assumed that this rate reduced the amount of viral aerosols accumulated in the classroom over an hour by half relative to having no ventilation \citep{american2002ansi}. 

We evaluated the worst case, where a poorly ventilated room  had 0 ACH and the risk of aerosol transmission was not reduced at all by ventilation. In addition, we evaluated the intervention where ventilation was improved to 3 ACH, which reduced the overall dose of transmission from aerosols by a factor of 4 relative to no ventilation.





\subsubsection{Class Type:}
The type of class determines the intensity of respiratory activity that occurs in the room, which corresponds to different rates of viral aerosol emission. 
We assumed breathing (without other respiratory activities) was the dominant type of respiratory activity for students attending lectures and that the effect of occasional speaking (e.g., asking and answering questions) was negligible. However, our simulation is also able to handle activities such as talking and singing. 

\subsection{Risk Over the Semester}

Given a set of interventions, we adopted the following procedure to assess the risk of transmission for students and instructors across the entire semester. More details are given in Appendix A2. 
We only considered undergraduate students, but our results easily translate to graduate or professional students who typically take fewer classes. 

We first used our classroom simulation tool to estimate the risk of transmission per hour spent in the classroom $\eta$, conditioned on the class having an infectious source case. 
Given average class size $n_0$ and prevalence of infectious individuals at the University $p$, the probability that a susceptible student attends a class with an infectious student is given by
$$1-(1-p)^{n_0-1}\approx (n_0-1)\cdot p.$$

This approximation is justified since $p$ is small at the University under our projections for the Fall 2021 semester (we also impose a prior on $p$ to incorporate uncertainty in its estimated value). 
Multiplying the above expression by $\eta$ gives the unconditional probability of infection per hour of class. Assuming a student's average time spent in class per semester to be $T$ hours, the probability of infection in class over a semester was given by
$$1-(1-\eta\cdot(n_0-1)\cdot p)^T \approx \eta\cdot (n_0-1)\cdot p\cdot T.$$
\editedit{The linearization approximations above produce upper bounds for the actual quantities. We erred on the conservative side and adopted these approximations for downstream computation. Appendix A1 further discusses the magnitude of the approximation errors.}

Lastly, we generated a distributional estimate using 100,000 samples, with each sample representing the semester-wise risk associated with a specific parameter configuration drawn from the priors.

A similar procedure was applied to faculty and graduate student instructors. We assumed that the instructor is sufficiently distanced from the students so that risk only arises from transmission over long distances. We adjusted the unconditional probability of infection per hour of class to account for their population sizes relative to the undergraduate population size. We also assumed the $T$ hours of classes are divided between faculty and graduate student instructors with a 2:1 ratio.

\section{Results and Recommendation for the Fall 2021 Semester}

\begin{figure}
    \centering
    \caption{Average number of secondary infections over one hour of lecture with one positive student for different intervention settings, assuming 90\% vaccination rate. Blue and orange represent unrestricted and fixed seating respectively. Solid and dashed lines represent 0\% and 100\% masking respectively.}
    \includegraphics[width=0.95\textwidth]{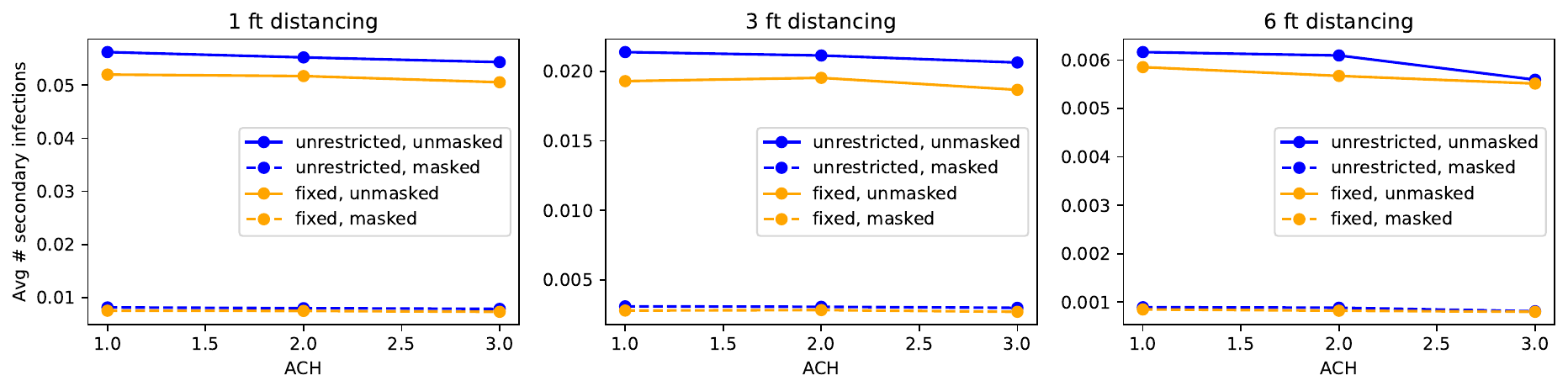}
    \label{fig:sim_results_all_configs}
\end{figure}

\editedit{
Here, we summarize the results and recommendations of our modeling analysis. Table~\ref{tab:single_classroom_results} and Figure~\ref{fig:sim_results_all_configs} present the expected number of secondary infections during a one-hour lecture with one positive student under different intervention combinations, assuming a 90\% vaccination rate. 
Our model results showed that different seating policies as well as different ventilation conditions both resulted in comparable risk. 
Even though increasing distancing provided a large risk reduction, such benefit was deemed to be outweighed by logistical difficulty as well as reduction in class capacity. 
On the other hand, masking was much more effective than enforcing a fixed seating plan and increasing ventilation. 
Indeed, requiring masking in dense classrooms with unrestricted seating was easy to implement, incurred little logistical or financial cost, and allowed classes to be held at full capacity. Our following analysis shows that this combination of interventions resulted in acceptable risk over the semester.
}

\begin{table}[]
\centering
\caption{Average number of secondary infections over one hour of lecture with one positive student for different intervention settings, assuming 90\% vaccination rate.}
\begin{tabular}{|l|l|ll|ll|}
\hline
 &  & \multicolumn{2}{l|}{unrestricted seating} & \multicolumn{2}{l|}{fixed seating} \\ \hline
 &  & \multicolumn{1}{l|}{unmasked} & masked & \multicolumn{1}{l|}{unmasked} & masked \\ \hline
\multirow{3}{*}{1 ft distancing} & 1 ACH & \multicolumn{1}{l|}{5.62E-2} & 8.12E-3 & \multicolumn{1}{l|}{5.20E-2} & 7.54E-3 \\ \cline{2-6} 
 & 2 ACH & \multicolumn{1}{l|}{5.52E-2} & 8.00E-3 & \multicolumn{1}{l|}{5.17E-2} & 7.45E-3 \\ \cline{2-6} 
 & 3 ACH & \multicolumn{1}{l|}{5.43E-2} & 7.88E-3 & \multicolumn{1}{l|}{5.05E-2} & 7.33E-3 \\ \hline
\multirow{3}{*}{3 ft distancing} & 1 ACH & \multicolumn{1}{l|}{2.14E-2} & 3.10E-3 & \multicolumn{1}{l|}{1.93E-2} & 2.80E-3 \\ \cline{2-6} 
 & 2 ACH & \multicolumn{1}{l|}{2.11E-2} & 3.07E-3 & \multicolumn{1}{l|}{1.95E-2} & 2.83E-3 \\ \cline{2-6} 
 & 3 ACH & \multicolumn{1}{l|}{2.06E-2} & 2.99E-3 & \multicolumn{1}{l|}{1.87E-2} & 2.71E-3 \\ \hline
\multirow{3}{*}{6 ft distancing} & 1 ACH & \multicolumn{1}{l|}{6.17E-3} & 8.94E-4 & \multicolumn{1}{l|}{5.86E-3} & 8.49E-4 \\ \cline{2-6} 
 & 2 ACH & \multicolumn{1}{l|}{6.10E-3} & 8.84E-4 & \multicolumn{1}{l|}{5.67E-3} & 8.23E-4 \\ \cline{2-6} 
 & 3 ACH & \multicolumn{1}{l|}{5.60E-3} & 8.11E-4 & \multicolumn{1}{l|}{5.52E-3} & 8.00E-4 \\ \hline
\end{tabular}
\label{tab:single_classroom_results}
\end{table}

We next present the simulated distributions of infection risk for students and instructors. For students, we focused on undergraduates and assumed a total of 15,000. For instructors, we accounted for both faculty and graduate student instructors, with estimates of 850 faculty and 3,120 graduate student instructors (details are given in Appendix A2). 
\editedit{Our estimated risk for undergraduates can be thought of as a representative upper bound for all students, including graduate and professional ones. At our university, graduate students either take a similar number of classes as undergraduates (e.g., one-year masters and MBA students), or fewer classes (e.g., two-year masters and early-stage PhD students), if not none (late-stage PhD students).
A small number of these classes may be shared by undergraduate and graduate students. As most infection spikes had occurred among undergraduates and the prevalence among graduate students was usually lower, the risk for undergraduate constituted a conservative estimate of the general risk for all students.
}

\subsubsection{Students Classroom Risk:}

\begin{figure}[h!]
    \centering
    \caption{Distribution of risk of lecture transmission for a student across the entire Fall 2021 semester, over $10^5$ simulation trials.
    } 
    \includegraphics[width=0.7\textwidth]{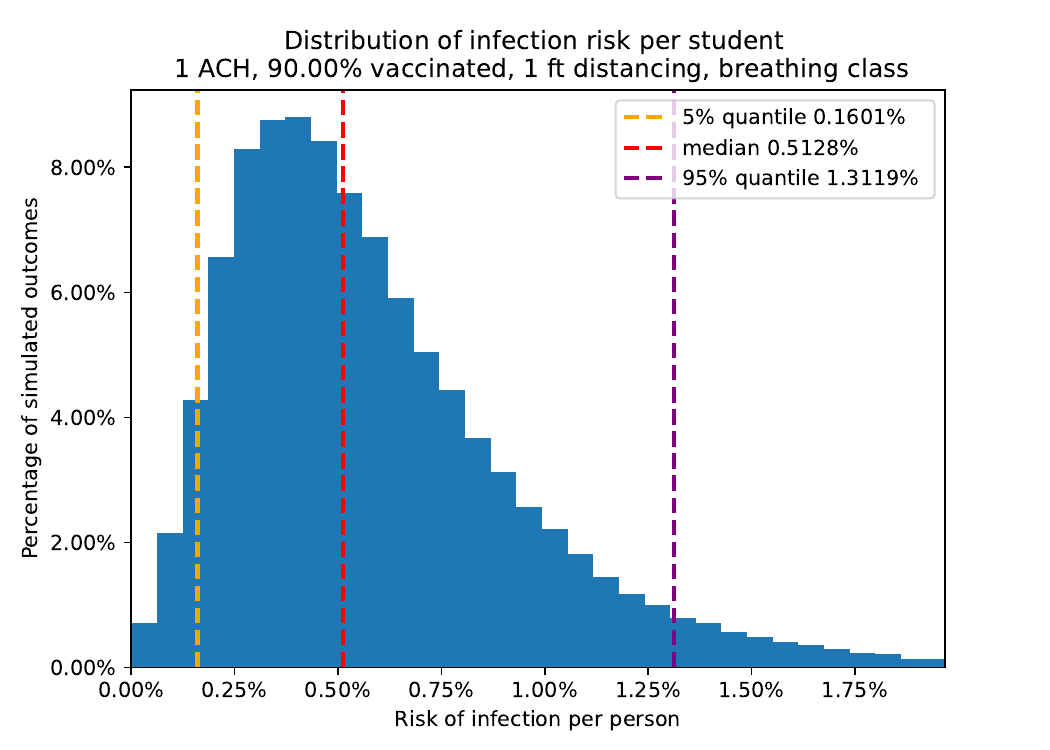}
    \label{studentlecturerisk.fig}

    {\small\textit{Note}: Median risk (red dashed line) is 0.51\%; the 5\% and 95\% quantiles (orange and purple dashed lines) are 0.16\% and 1.31\% respectively. }
\end{figure}

We projected that the median risk of infection per student due to lecture transmission in Fall 2021 would be 0.5\% at 90\% vaccination rate. (An earlier version of this analysis \citep{fall2021_classroom_modeling_report} predicted this number to be 0.4\% due to outdated parameters.) Figure \ref{studentlecturerisk.fig} shows the estimated distribution of this risk across 100,000 simulated outcomes; the median is indicated by the red dashed line. The right tail of the estimated risk distribution mainly results from the right tail of the log-normal prior over the prevalence parameter.

\subsubsection{Instructor Classroom Risk:}

\begin{figure}[h!]
    \centering
    \caption{Distribution of risk of lecture transmission for a vaccinated faculty instructor across the entire Fall 2021 semester, over $10^5$ simulation trials.}
    \includegraphics[width=0.7\textwidth]{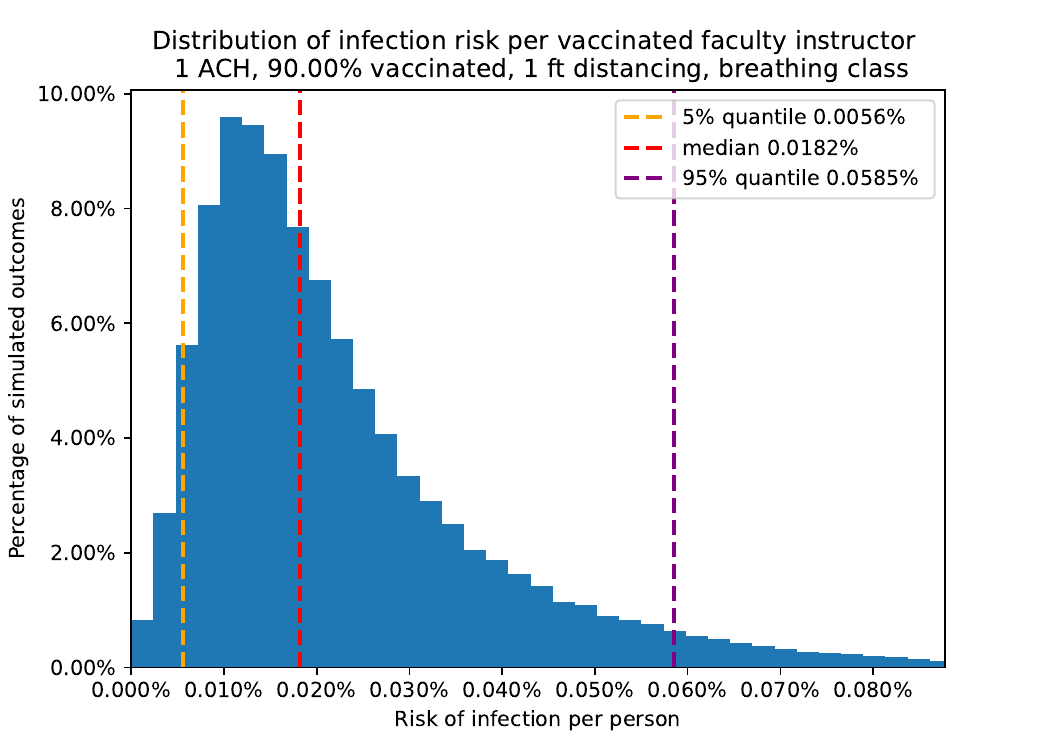}
    \label{facultylecturerisk.fig}

    {\small\textit{Note}: Median risk (red dashed line) is 0.018\%; the 5\% and 95\% quantiles (orange and purple dashed lines) are 0.0056\% and 0.059\% respectively. }
\end{figure}

\begin{figure}[h!]
    \centering
    \caption{Distribution of risk of lecture transmission for a vaccinated graduate student instructor across the entire Fall 2021 semester, over $10^5$ simulation trials.} 
    \includegraphics[width=0.7\textwidth]{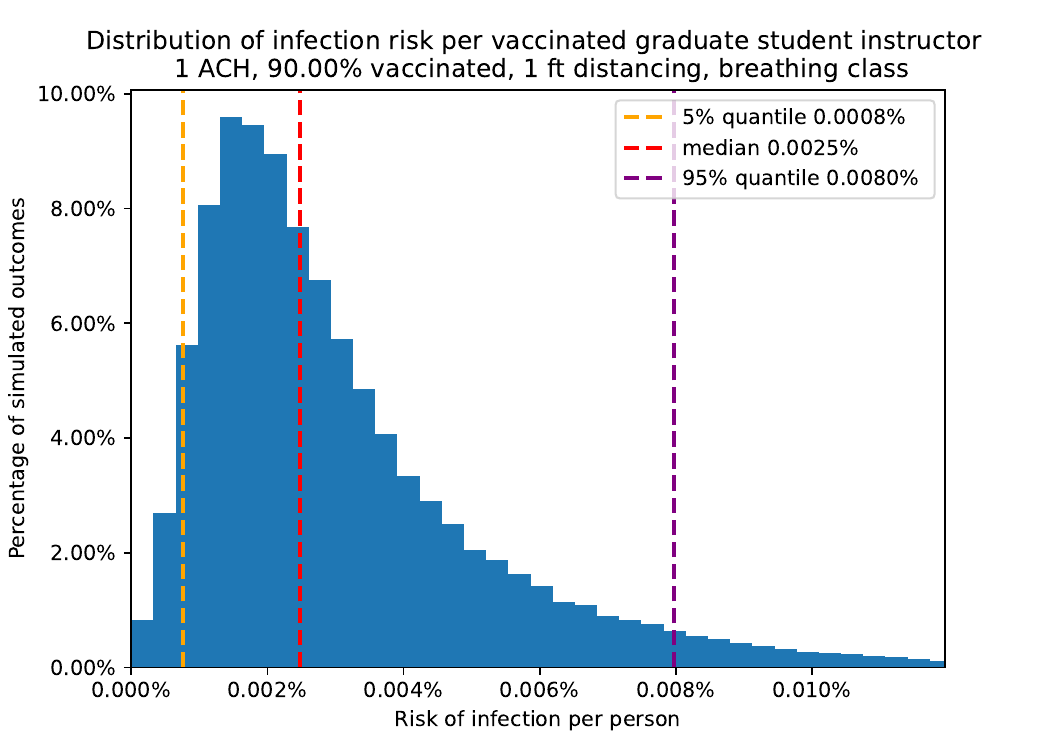}
    \label{gradstudentlecturerisk.fig}

    {\small\textit{Note}: Median risk (red dashed line) is 0.0025\%; the 5\% and 95\% quantiles (orange and purple dashed lines) are 0.0008\% and 0.008\% respectively. }
\end{figure}

We projected that the median risk of infection per instructor due to lecture transmission in Fall 2021 would be 0.02\% for vaccinated faculty instructors and 0.003\% for vaccinated graduate student instructors. The estimated distribution of risk across simulated outcomes is presented in Figures \ref{facultylecturerisk.fig} and \ref{gradstudentlecturerisk.fig}. 

The risk was approximately doubled for an unvaccinated instructor. Since over 99\% of professorial faculty had been vaccinated by the start of the semester \citep{dailysun_fall2021}, and those who chose not to be vaccinated would be highly cautious, we only show the estimated risk for the vaccinated instructors here.

The projected risk for instructors was much lower than that for students. This is mainly due to the modeling choice that instructors were not subject to short-distance transmission, based on the natural distancing between instructors and students in classrooms. In addition, instructors spent less time in class over a semester compared to students.

\subsection{Recommendations}

Based on our analysis, we believed that fully dense in-person classes, with masking enforced, could be safely implemented for the Fall 2021 semester. We estimated the total risk of classroom transmission per student across the entire semester to be around 0.5\%, or roughly 1 in 200. For faculty and graduate students, the estimated risk of classroom transmission was even lower, roughly 1 in 5,000 to 40,000 across the entire semester. An individual's odds of being struck by lightning in their life is on the order of 1 in 10,000 \citep{NWS_2019}, which is comparable.

Under the assumption that 15,000 students would return to the University for the Fall 2021 semester, we conservatively anticipated an additional 75 cases due to classroom transmission, \editedit{with this figure rising to 119 under 90\% masking compliance}. We did not expect these additional cases to strain the testing and quarantine capacity of the university; the University's testing infrastructure was able to handle tens of thousands of tests per week and the university had the capacity to quarantine hundreds of students at a time. In addition, given the estimate of COVID-19 hospitalization rates for college-age students of 0.005\% \citep{covidnet}, we did not expect that any students would be hospitalized from an infection due to classroom transmission. Finally, assuming 850 faculty members and 3,120 graduate students serve as instructors in the Fall 2021 semester, we did not expect to observe any instructor cases linked to classroom transmission.

\section{Evaluation}

To evaluate our modeling framework and recommendations, we retrospectively investigated COVID-19 cases from August 26, 2021 to December 7th, 2021. We excluded a spike in cases due to importation of the Omicron variant in mid December for two reasons: it happened after the end of the instruction period when no classes were in session, and our modeling analyses and recommendations were specific to the Delta variant.

\subsection{Student Transmission}

We present the following body of evidence that minimal classroom transmission occurred among students during the Fall 2021 semester \citep{track_transmission_webarchive}.

\begin{figure}[h!]
    \centering
    \caption{Daily new cases among undergraduate students during the Fall 2021 semester.}
    \includegraphics[width = \textwidth]{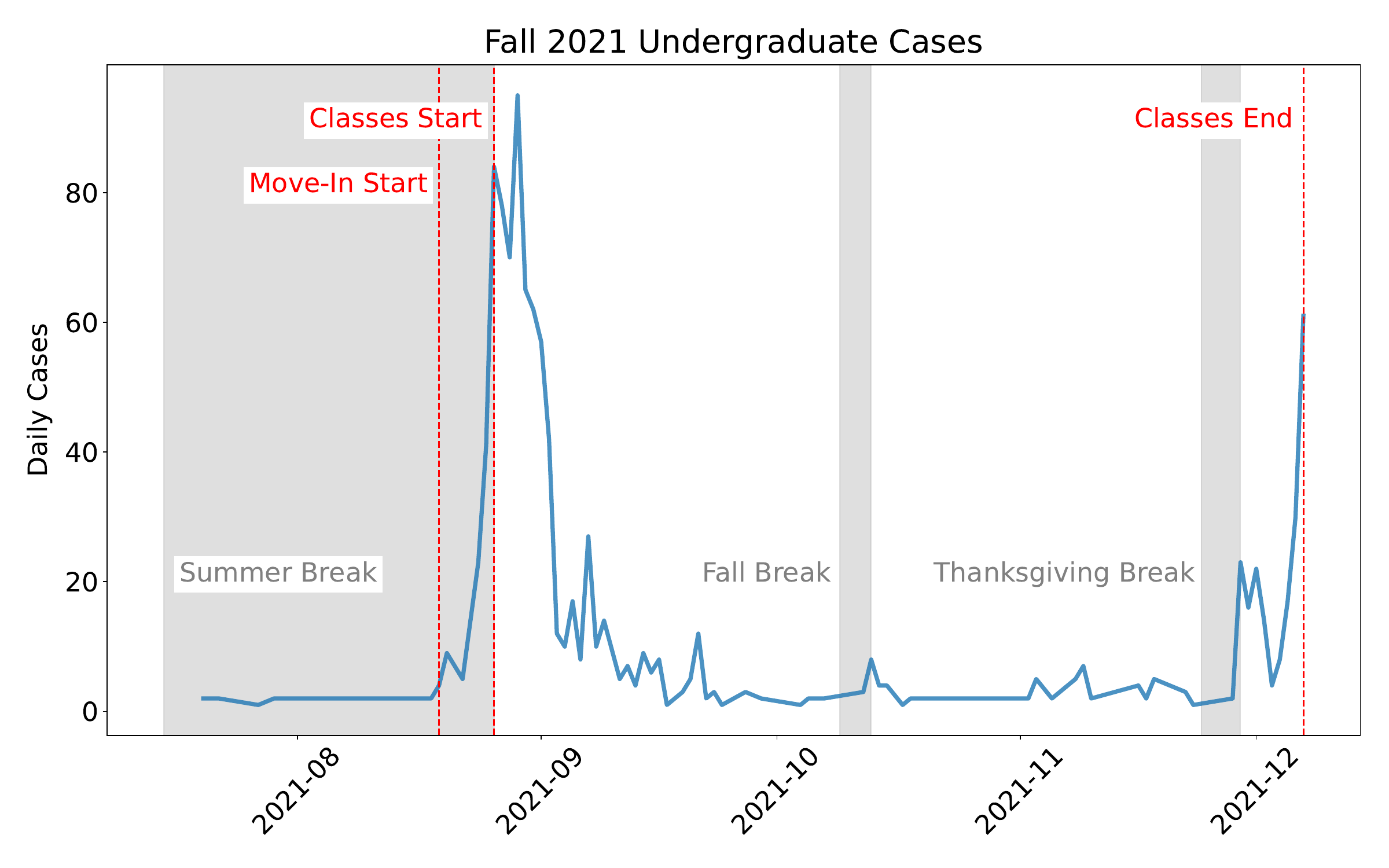}
    \label{timeseries_cases.fig}
    {\small\textit{Notes}: The gray intervals indicate school breaks (classes not in session). The largest spikes in cases occurred after events that involve significant student travel (move-in period and Thanksgiving break). December 7th was the last day of classes for the semester. The Omicron variant was responsible for the rightmost spike in cases, which occurred after classes ended.}
\end{figure}

\begin{enumerate}
    \item When a student tested positive during the Fall 2021 semester, the University tested all students attending the same class to the extent feasible. In addition, the genetic sequencing of positive cases were compared to determine if cases were related. These investigations did not yield evidence of classroom transmission.
    
    \item We collected seating data for a class held in a lecture hall that contained over 1000 students. When a student in that class tested positive, we investigated to see if any students seated near the infected student subsequently tested positive. While this data is sparse, there were 20 instances of an infected student sitting within 3 seats of susceptible students. None of these cases was associated with a susceptible student testing positive. 
    
    \item Throughout the semester, the weeks with the highest rates of on-campus transmission corresponded to breaks when classes were not held. This is consistent with travel and social gatherings, rather than classes, driving COVID-19 transmission on campus, as was also observed in previous semesters \citep{feb2021_report}. Figure \ref{timeseries_cases.fig} shows the daily count of new cases for undergraduate students. The outbreaks occurred right after students returned to campus from breaks.
    
    \item Contact tracing revealed that most positive cases can be linked by social gatherings, cohabitation, or travel.
    
\end{enumerate}
This collection of evidence strongly suggests that classroom transmission was rare during the Delta wave of the Fall 2021 semester at the University.

\subsection{Instructor Transmission}
Throughout the Fall 2021 semester, only a single faculty member tested positive for COVID-19 at the University. In addition, the prevalence of positive cases among graduate students was 4 times lower than the prevalence among undergraduate students. In all, infection rates among faculty and graduate students were much lower compared to the rest of the University population, which suggests that in-person teaching did not appreciably increase the risk of contracting COVID-19 during the Fall 2021 semester relative to other sources of transmission.

\section{Extensions}

\subsection{Beyond the Classroom}



While the main focus of our work was to model the risk of COVID-19 transmission in lectures and classrooms, we received many requests from the University administration to assess the risk of holding extra-curricular events and gatherings during the Fall 2021 semester. We were able to modify our modeling framework to accommodate these requests; we used the same model structure but updated our parameters to model eating, singing, socializing, and other events that occur in social gatherings. Our modeling analysis influenced the following decisions. For Homecoming weekend, we determined that the homecoming football game and the Class of 2020's belated graduation ceremony were relatively low-risk events. However, we found that parties and festivities that occur after formal events incur substantially higher risk of COVID-19 transmission. As a result, we recommended canceling post-homecoming festivities such as the fireworks and light shows on-campus; these recommendations were accepted by the administration. We did not observe a large spike in cases on campus after homecoming weekend.

We were also asked by executive-level decision-makers at the University to assess the risk of the campus-sponsored Rosh Hashanah dinner. We determined that given the high vaccination rate on campus, this event would be safe to attend. Finally, we were asked to evaluate the risk of indoor physical education classes and music and choir classes. We modified our model to account for the increased aerosol emission due to these activities and determined that it was safe to hold these classes with dense seating configurations. No cases were linked to these courses at the end of the semester.

The flexibility of our framework in accommodating these \textit{ad hoc} situations indicates that our framework can be applied to other industries besides higher education, to plan indoor space use to avoid the transmission of infectious diseases across a range of applications.



\subsection{Beyond COVID-19}

 In addition, by re-fitting the models on short range and long range transmission and by re-estimating the parameters on vaccine and mask efficacy, we can easily adapt our framework to model other respiratory diseases or COVID-19 variants. As such, our modeling framework can be used to assess infection risk in future pandemics across various settings.


\subsection{Conclusion}

Our modeling framework for COVID-19 transmission in classrooms allowed the University to analyze the risk of holding in-person classes and compare the effectiveness of interventions. Using the recommendations provided by our modeling framework, the University was able to return to pre-pandemic levels of in-person instructions for the Fall 2021 semester, improving the educational experience of students compared to previous semesters while ensuring safety. Post-hoc analysis at the end of the semester confirmed that classroom transmission was rare and that teaching in-person classes was a low-risk activity. Finally, our modeling framework is flexible and can be adapted to model infections risk for respiratory diseases across a wide range of applications.




\ACKNOWLEDGMENT{%
This work was conducted under the support from Cornell University when the authors served in the Cornell COVID mathematical modeling team.
This work was partially supported by National Science Foundation grant DMS2230023. 
}


\bibliographystyle{informs2014} 
\bibliography{references} 


%
%
%
\newpage
\pagenumbering{arabic}

\begin{APPENDIX}{Modeling and Simulation Details}

We describe the details of our modeling and simulation for estimating secondary infections in classrooms and projecting to the risk over the entire semester. In Section A1, we introduce the mechanism behind our simulation tool. In Section A2, we state the assumptions and parameter estimates in different components of the simulation. In Section A3, we develop the mathematical model for the transmission probability between a source case and a susceptible person over different distances, which is an important component in the simulation. 
Throughout the appendix, we use \texttt{\#} to denote ``number of" and \texttt{\%} to denote ``fraction of".

\section{A1. Simulation}

We implement a simulation tool in Python to simulate classrooms under different scenarios. We investigate the effect of different intervention measures (masking, increasing distancing, ventilation) on transmission risk under varying simulation parameters (vaccination rate among students, vaccine efficacy). We describe the simulation mechanism in this section. In Section A2, we discuss the assumptions and values for these parameters in more detail. 

Our simulation has two stages. In the first stage, we estimate the conditional probability that a student or instructor is infected in a one-hour class given there is one positive source case in the same class. In the second stage, we extrapolate the probability that a student or instructor becomes infected due to attending or teaching classes over the entire semester.

We omit the scenarios where there are two or more source cases in the same classroom at the same time. Such scenarios are relatively unlikely because prevalence is low. 
Moreover, the fact that a few positives occur in the same classrooms at the same time makes reality slightly more optimistic than our estimates: in our estimates, everyone else in the classroom is susceptible while in reality, the other infectious individual cannot be infected again.
In addition, the increase in the risk of infection created by adding a second positive is smaller than the increase created by adding a first positive\footnote{
This follows (1) when the dose resulting from each positive is independent and identically distributed and (2) from concavity of the probability of infection as a function of the dose, given in Appendix A3. In particular, let $V_1$ and $V_2$ be the (strictly positive) dose to a given susceptible person associated with the first and second positives in a classroom, so that the dose is $V_1$ if there is one positive in the classroom and $V_1 + V_2$ if there are two. We assume that $V_1$ and $V_2$ are independent and identically distributed after marginalizing over the random locations of the two positive individuals. Then let $P(v)$ be the probability of infection given a dose $v$, as given in Section A3.  Since $P$ is concave, and also using that $P(0)=0$, then for strictly positive $V_1$ and $V_2$, $P(V_1 + V_2) - P(V_1) \leq P(V_2) - P(0)$. Then, because $V_1$ and $V_2$ are identically distributed, $\mathbb{E}[P(V_1 + V_2) - P(V_1)] \leq \mathbb{E}[P(V_2) - P(0)] = \mathbb{E}[P(V_2)] = \mathbb{E}[P(V_1)]$. The left-hand side $\mathbb{E}[P(V_1 + V_2) - P(V_1)]$ is the increase in the risk of infection created by adding a second positive, and the right-hand side $\mathbb{E}[P(V_1)]$ is the increase in risk from adding the first positive.
}. The level of optimism introduced by this fact is extremely small, however, and our assumption produces nearly the same estimate as one that allows multiple positives to be in the same classroom. 

\subsection{Generating Classroom Seating Arrangements}

We simulate seating at different density levels by assigning a fixed number of individuals to classrooms of different sizes. We assume that an average class contains $n_0$ = 50 students and one instructor. From university floor plans, we identify three representative rooms, namely Hollister 206, Gates G01, and Rockefeller 201, that correspond to roughly 1', 3', and 6' distancing respectively for 50 students. The corresponding seating capacities are presented in Table~\ref{tab:seating_capacity}.

\begin{table}[h]
\caption{Seating capacity for Hollister 206, Gates G01, and Rockefeller 201.}
\begin{tabular}{|l|l|l|}
\hline
Room & Pre-COVID capacity (1' distancing) & COVID capacity (6' distancing) \\ \hline
Hollister 206 & 52 & 12 \\ \hline
Gates G01 & 156 & 22 \\ \hline
Rockefeller 201 & 383 & 56 \\ \hline
\end{tabular}\label{tab:seating_capacity}
\end{table}

Using the seating plan tool developed by the Committee on Teaching Reactivation Options (C-TRO) team \citepAP{greenberg2021automated}, we identify seats in these classrooms and assign maximally distanced seats such that approximately 50 students can fit in each room. The generated seating plans are displayed in Figure~\ref{fig:classroom_seating_plans}. The rooms used have extra space above and beyond what is required to accommodate the social distancing requirements we have assumed. This extra room does not offer additional benefit in our simulations as it is not used. We assume the instructor is standing in the front of the classroom, with at least 6 feet distance from all the students. 

\begin{figure}
    \centering
    \caption{Seating plans generated for Hollister 206, Gates G01, and Rockefeller 201. They correspond to 1', 3', and 6' social distancing, respectively. The green dots represent available seats (i.e., seats that are allowed to be occupied) and the empty circles are considered unavailable in our simulation tool.}
    \includegraphics[width=0.85\textwidth]{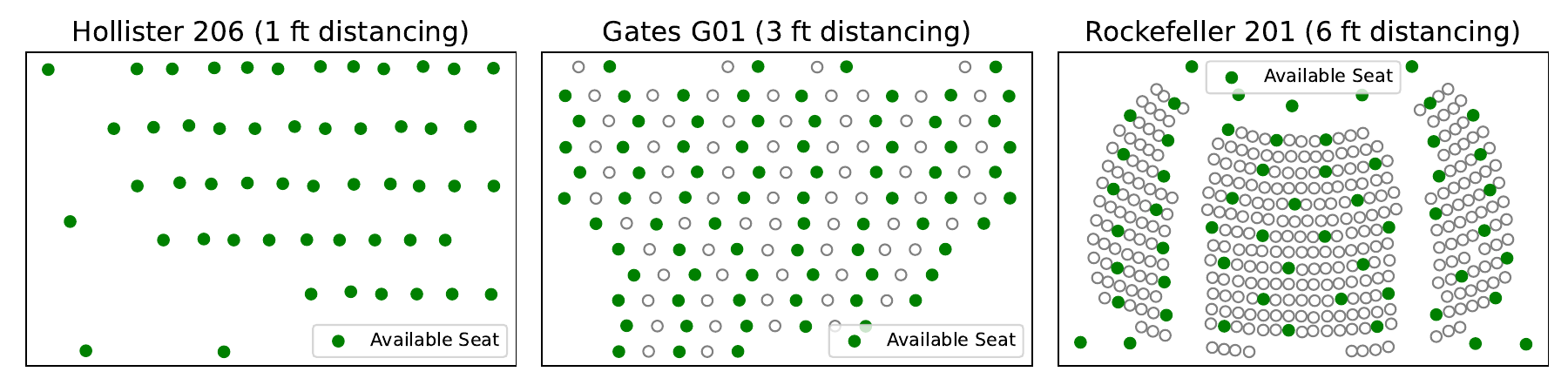}
    \label{fig:classroom_seating_plans}
\end{figure}

We initially consider two seating policies: (1) fixed seating, where students are randomly assigned to their own seats independent of their vaccination statuses; (2) unrestricted seating, where students can freely choose among the allowed seats. For unrestricted seating, we pessimistically assume that unvaccinated students tend to sit together, as the same demographic factors that cause students to be unvaccinated upon arrival to Ithaca may also create social connections.
Since fixed seating is operationally difficult to implement, and initial simulations show that these two seating policies lead to comparable risk (Figure~\ref{fig:seating_policies_comparison})\footnote{\editedit{Figure~\ref{fig:seating_policies_comparison} was generated at the project's outset using outdated parameters. Back then, we did not have a full understanding of masking efficacy, so we assumed a 20\% reduction in risk due to masking.}}, the University decided to adopt unrestricted seating. 
Figure~\ref{fig:classroom_seating_plans_vax_unvax} shows examples of unrestricted seating arrangements at different distancing levels. 

\begin{figure}
    \centering
    \caption{Initial comparison of fixed and unrestricted seating policies. We show the average number of student secondary infections over a one-hour lecture with a positive student for different vaccination rates (from 40\% to 100\%), distancing levels (1, 3, and 6 feet distancing), and masking rates (0\% and 100\%). The differences between the solid lines (unrestricted seating) and dashed lines (fixed seating) are small and further decrease as vaccination rate increases. The vaccination rate we assume in our subsequent simulations, 90\%, is indicated by the grey dotted line. \editedit{Here, masking is assumed to provide a 20\% reduction in risk; this would be later updated to roughly range from 75\% to 96\% (Table~\ref{tab:sim_params}).}}
    \includegraphics[width=0.85\textwidth]{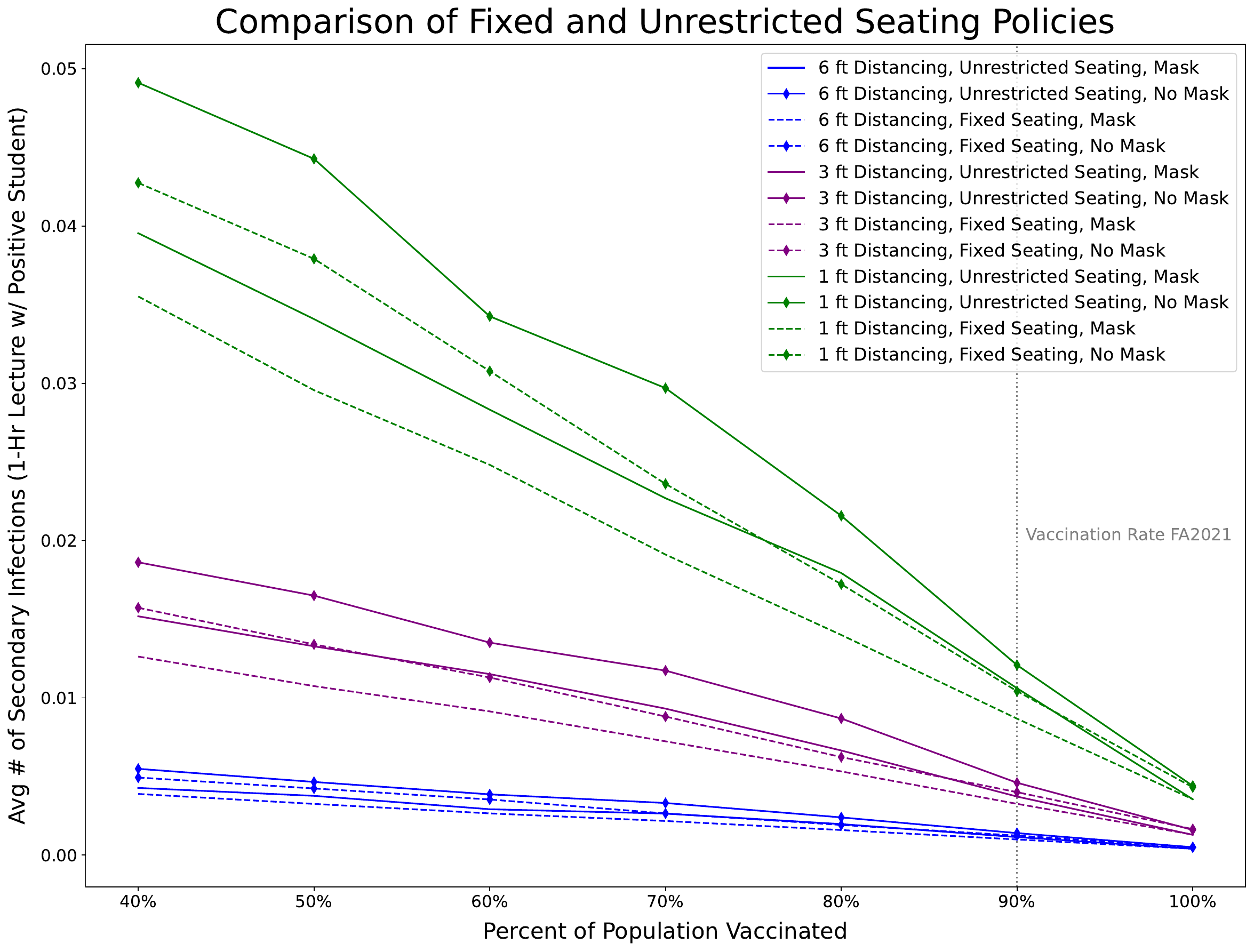}
    \label{fig:seating_policies_comparison}
\end{figure}

\begin{figure}
    \centering
    \caption{Example simulated seating arrangement of 50 students at different levels of social distancing. 
    At each distancing level, students are placed into available seats shown in Figure~\ref{fig:classroom_seating_plans}.
    The red dots represent unvaccinated students, the blue dots represent vaccinated students, and the empty circles represent unavailable seats due to the distancing requirement or empty available seats. 
    We assume pessimistically that unvaccinated students tend to sit together.}
    \includegraphics[width=0.85\textwidth]{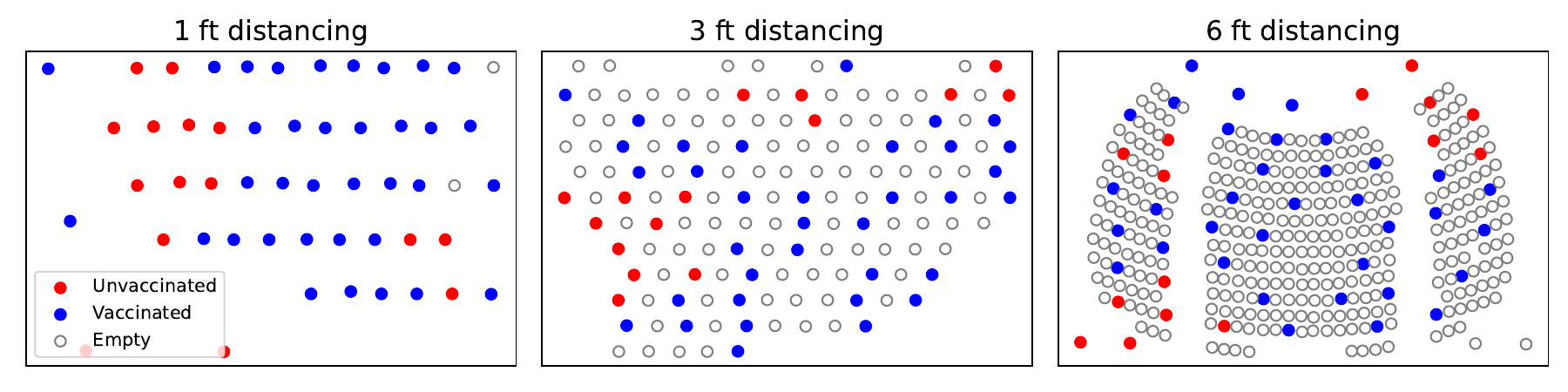}
    \label{fig:classroom_seating_plans_vax_unvax}
\end{figure}

\subsection{Stage 1: Simulating Infections in a Single Classroom}


In the first stage, we estimate the conditional probability that a student or instructor is infected in a one-hour class given there is one positive source case in the same class. We do this by simulating the expected number of secondary infections that occur in a one-hour class with $n_0=50$ students, where initially a single student is infected. The simulation depends on the following configuration parameters: seating density, percentage of students vaccinated, and vaccine efficacy (\texttt{VE}).
In particular, vaccine efficacy $\texttt{VE}=(v_{\text{source}}, v_{\text{susceptible}})$ is a tuple of two parameters characterizing the reduction in infectivity of a vaccinated source case and the reduction in infection probability of a vaccinated susceptible person respectively. 
We assume each student is vaccinated independently with probability equal to the vaccination rate among the students.

In this stage, we assume everyone is unmasked. This allows the flexibility to model varying masking rate among students as well as uncertainty in masking effectiveness in the next stage of the simulation.

We loop over all the possible combinations of seating density and $\texttt{VE}$ (values of \texttt{VE} are discussed in detail in Section A2). For each configuration with fixed parameter values, one run of the simulation proceeds in the following steps:
\begin{itemize}
    \item Simulate a seating arrangement of $n_0=50$ students in the corresponding classroom.
    \item Choose one student uniformly at random as the source case and simulate their vaccination status.
    \item For each susceptible student, first simulate their vaccination status, then compute the probability $p$ that they are infected over the 1-hour lecture depending on their relative location to the source case. The mathematical model that produces this probability is derived in Appendix A3. 
    \item Return the sum of the individual infection probabilities in the last step divided by $(n_0-1)$. This is the average probability that a susceptible student becomes infected in a one-hour class given that there is an infectious student in the class, denoted $\eta_{\text{student}}(\text{density}, \texttt{VE})$.
    \item We also compute the infection probability for the instructor, who is assumed to be subject to only the risk of aerosol transmission due to sufficient distancing, denoted $\eta_{\text{instructor}}(\text{density}, \texttt{VE})$.
\end{itemize}
For each simulation configuration, we perform 500 replications to obtain the average conditional probability of infection over 1 hour given a positive case in the classroom. 

\subsection{Stage 2: Extrapolating Classroom Transmission Risk over the Semester}

In the second stage, we extrapolate the probability of infection over the entire semester based on results in Stage 1 and additional parameters (prevalence, masking effectiveness, and fraction of population masked). 

To deal with uncertainty in the parameter values, 
we impose independent prior distributions on each of them. We sample parameter configurations from the prior and compute the output at each configuration, which produces a distribution of the output. In particular, for each sampled parameter configuration (vaccine efficacy $\texttt{VE}= (v_{\text{source}}, v_{\text{susceptible}})$, masking effectiveness $m$, prevalence $p$) and a selected seating density, we perform the following computation for an average student:
\begin{itemize}
    \item Get $\eta_{\text{student}}(\text{density}, \texttt{VE})$ for the seating density and sampled vaccine efficacy from Stage 1 simulation results. 
    \item Adjust for masking according to $\beta_{\text{masked}}$, the fraction of students wearing a mask, and masking effectiveness $m$, the reduction in transmission probability due to mask-wearing. We obtain the transmission probability adjusted for masking, conditioned on the existence of an infectious student in the class (we omit the terms (density, \texttt{VE}) for brevity):
    
    \begin{equation}
        \tilde{\eta}_{\text{student}}=\eta_{\text{student}}(\text{density}, \texttt{VE}) \cdot [\beta_{\text{masked}}\cdot  (1-m) + (1 - \beta_{\text{masked}})].\label{eq:prob_inf_masking}
    \end{equation}
    
    This expression is an approximation in that we set $m$ to be the reduction in transmission probability when both the source and susceptible individuals are masked, and we are not explicitly modeling the case where only one of them is masked. In practice, we assume $\beta_{\text{masked}}=1$ in our modeling since masking compliance in previous semesters was very high \citepAP{faculty_townhall}.
    
    \item Under the assumption that infectious students are uniformly distributed across classes, the probability that a susceptible student attends class with an infectious student is given by
    \begin{equation}
        1-(1-p)^{n_0-1}\approx (n_0-1)\cdot p, \label{eq:prob_inf_student_class}
    \end{equation}
    where the approximation is justified since prevalence, $p$, is small in practice; \editedit{see below for further discussion of the approximation}. Thus, the unconditional infection probability for a student over a one-hour lecture is given by
    $$\tilde{\eta}_{\text{student}}\cdot (n_0-1)\cdot p.$$

    \item Extrapolate the infection probability over the semester. 
    Suppose an undergraduate student spends on average $\tau_{\text{UG}}$ hours in class in a semester.
    The probability of infection over $\tau_{\text{UG}}$ hours is
    \begin{equation}
        \text{Risk}_{\text{student}} = 1-(1-\tilde{\eta}_{\text{student}}\cdot (n_0-1)\cdot p)^{\tau_{\text{UG}}} \approx \tilde{\eta}_{\text{student}}\cdot (n_0-1)\cdot p \cdot \tau_{\text{UG}}.\label{eq:p_inf_student_semester}
    \end{equation}
\end{itemize}


The above procedure does not eliminate already-infected students from the susceptible pool over time, so the estimated results are more pessimistic than reality. A similar computation can be done for faculty and graduate student instructors. 
\begin{itemize}
    \item We adjust for masking for instructors in a similar way as for students:
    $$\tilde{\eta}_{\text{instructor}}=\eta_{\text{instructor}}(\text{density}, \texttt{VE}) \cdot [\beta_{\text{masked}}\cdot  (1-m) + (1 - \beta_{\text{masked}})].$$
    \item We assume faculty instructors teach a fraction $\beta_{\text{faculty}}$ of all class hours, while graduate students teach the rest. 
    Let $n_{\text{UG}}$, $n_{\text{faculty}}$, and $n_{\text{graduate}}$ denote the number of undergraduates, faculty instructors, and graduate student instructors, respectively. 
    The average hours a faculty instructor spends in class teaching in a semester is
    $$\tau_{\text{faculty}}=\frac{n_{\text{UG}} \cdot \tau_{\text{UG}} \cdot \beta_{\text{faculty}}} {n_0\cdot n_{\text{faculty}}},$$
    where ${n_{\text{UG}} \cdot \tau_{\text{UG}}} / {n_0}$ gives the total undergraduate lecture hours to be taught by all instructors. 
    Similarly, the average hours a graduate student instructor spends in class teaching in a semester is $$\tau_{\text{graduate}}=\frac{n_{\text{UG}} \cdot \tau_{\text{UG}} \cdot (1-\beta_{\text{faculty}})} {n_0\cdot n_{\text{graduate}}}.$$
    \item With the same reasoning as in Equation~\ref{eq:p_inf_student_semester}, we derive the risk for instructors over the semester: 
    \begin{align*}
        \text{Risk}_{\text{faculty}} &= 1-(1-\tilde{\eta}_{\text{instructor}}\cdot (1-(1-p)^{n_0}))^{\tau_{\text{faculty}}} \approx \tilde{\eta}_{\text{instructor}}\cdot n_0 \cdot p \cdot \tau_{\text{faculty}}.\\
        \text{Risk}_{\text{graduate}} &= 1-(1-\tilde{\eta}_{\text{instructor}}\cdot (1-(1-p)^{n_0}))^{\tau_{\text{graduate}}} \approx \tilde{\eta}_{\text{instructor}}\cdot n_0 \cdot p \cdot \tau_{\text{graduate}}.
    \end{align*}
\end{itemize}

In the next section we give details on how we obtain estimates for $\tau_{\text{UG}}$, $n_{\text{UG}}$, $n_{\text{graduate}}$, $n_{\text{faculty}}$, and $\beta_{\text{faculty}}$.

\editedit{
\paragraph{Quantification of approximation errors} For simplicity, we use a first-order Taylor approximation in Equation~\ref{eq:prob_inf_student_class} and Equation~\ref{eq:p_inf_student_semester}. We quantify the approximation errors here. 


For Equation~\ref{eq:prob_inf_student_class}, let $f(p) = 1-(1-p)^{n_0-1}$, so that $f'(p) = \frac{df}{dp} = (n_0-1)(1-p)^{n_0-2}$. The first-order Taylor expansion around 0 is given by $f(p) = f(0) + f'(0) p + R_1(p) = (n_0-1)p+R_1(p)$. By Taylor's Theorem, for some $c$ between 0 and $p$, the remainder $R_1(p) = \frac{f''(c)}{2} p^2$. The second derivative is $f''(c) = -(n_0-1)(n_0-2)(1-c)^{n_0-3}$, so $|f''(c)| \leq (n_0-1)(n_0-2)$.
In our modeling, the prior distribution for the prevalence on campus is LogNormal$(-6.157, 0.413)$ with median value $\exp(-6.157)=0.002$. Plugging in the median value for $p$, and noting that $n_0=50$, we derive the following bound for the magnitude of the remainder:
$$|R_1(p)| \leq \frac{(n_0-1)(n_0-2)}{2} \cdot 0.002^2 = 0.005.$$
Indeed, we can verify that $1-(1-0.002)^{49} = 0.093$ and the approximation $49\cdot0.002 = 0.098$, with a $5\%$ relative error. 

For Equation~\ref{eq:p_inf_student_semester}, $\tilde{\eta}_{\text{student}}=0.056$ for dense seating and 1 ACH without masking and other parameter values following Table~\ref{tab:sim_params}. 
By Taylor's Theorem, the bound for the magnitude of the remainder is 
$$
\frac{\tau_{\text{UG}}(\tau_{\text{UG}}-1)}{2} \cdot (\tilde{\eta}_{\text{student}}\cdot (n_0-1)\cdot p)^2 = 1.5.
$$
We can verify that $1-(1-0.056\cdot49\cdot0.002)^{315} = 0.825$ and $0.056\cdot49\cdot0.002\cdot{315} = 1.736$, so the absolute and relative approximation error is 0.911 and +105\% respectively. We acknowledge that the approximation provides a loose upper bound for the actual quantity being modeled, which is also directly computable, but we adopted the approximation values at the time of modeling. We were erring on the conservative side, producing estimates that were upper bounds of the true quantities being computed.
}



\section{A2. Assumptions and Parameters}


This section presents our assumptions and estimates for the parameters used in the modeling. A summary is given in Table~\ref{tab:sim_params}. For parameters with high uncertainty, we design sensible prior distributions for their values rather than using a point estimate. 
From the joint priors on the parameters, we sample $10^5$ parameter configurations and obtain a distribution for $\text{Risk}_{\text{student}}$, $\text{Risk}_{\text{faculty}}$, and $\text{Risk}_{\text{graduate}}$ respectively. 
We treat the median as a main point estimate and use the 5\% and 95\% quantiles as optimistic and pessimistic estimates. 
Running simulations at a large number of parameter configurations sampled from the priors enables a better understanding of how the possible outcomes are distributed. 


\begin{table}[htbp]
\caption{Simulation parameters.}\label{tab:sim_params}
\centering
\begin{tabular}{|l|l|}
\hline
\textbf{Parameter} & \textbf{Value / Prior distribution} \\ \hline
transmissibility of the Delta variant & 2.4 times that of Alpha \\ \hline
respiratory activity & breathing \\ \hline
$\beta_{\text{vaccinated}}$, fraction of students vaccinated & 90\% \\ \hline

$v_{\text{source}}$, 
reduction in a vac'd source's infectivity  &  see Table~\ref{tab:ve_params}\\ \hline
$v_{\text{susceptible}}$, 
reduction in a vac'd susceptible's infection prob.  & see Table~\ref{tab:ve_params} \\ \hline
$\beta_{\text{masked}}$, fraction of students masked & 100\% \\ \hline
$m$, masking effectiveness & $\mathcal{N}$(0.855, 0.0536), truncated to [0,1] \\ \hline
$p$, prevalence & LogNormal(-6.157, 0.413) \\ \hline
$\tau_{\text{UG}}$, avg hours an undergrad spends in class in a semester & 315\\ \hline
$n_0$, class size & 50\\ \hline
$n_{\text{UG}}$, number of undergraduates & 15,000\\ \hline
$n_{\text{faculty}}$, number of faculty instructors & 850\\ \hline
$n_{\text{graduate}}$, number of graduate instructors & 3,120\\ \hline
$\beta_{\text{faculty}}$, fraction of classes taught by faculty instructors & 2/3 \\ \hline
\end{tabular}
\end{table}

\begin{enumerate}
    \item \textbf{The Delta variant}
    \begin{enumerate}
        \item We assume the Delta variant had dominated all infections by the start of the Fall 2021 semester.
        \item The Delta variant is \textbf{2.4} times more transmissible than the non-variant strain. 
        The Alpha variant was approximately 50\% more transmissible than the original SARS-CoV-2 \citepAP{washington2021emergence}, and Delta is approximately 60\% more transmissible than Alpha \citepAP{callaway2021delta}. This gives a multiplicative increase of 1.5 * 1.6 = 2.4.
    \end{enumerate}
    
    \item \textbf{Type of respiratory activity}
    
    Respiratory activities of different intensities are associated with varying transmissibility. We assume breathing is the dominant type of respiratory activity for students attending lectures and that the effect of occasional speaking (e.g., asking and answering questions) is negligible. However, our simulation is able to handle activities such as talking and singing. 

    \item \textbf{Level of vaccination among undergraduates} ($\beta_{\text{vaccinated}}$)
    
    By the start of the Fall 2021 semester, 99\% of undergraduate students and professorial faculty were fully vaccinated \citepAP{dailysun_fall2021}. Erring on the conservative side, we used \textbf{90\%} for the fraction of vaccinated undergraduates at the time the analysis was performed. We compute the infection probability for vaccinated and unvaccinated instructors separately.

    \item \textbf{Vaccine efficacy} ($\texttt{VE}=(v_{\text{source}}, v_{\text{susceptible}})$)
    We base our estimates of vaccine efficacy on the literature available at the time of our modeling.
    \begin{enumerate}
        \item For $v_{\text{source}}$, we assume vaccination reduces the infectivity of a source case by \textbf{0}, \textbf{50\%}, and \textbf{71\%} with probabilities proportional to the sample size of the corresponding study \citepAP{levine2021initial, harris2021effect, brown2021outbreak}. See Table~\ref{tab:ve_params}.
        \item For $v_{\text{susceptible}}$, we assume vaccination reduces the risk of infection for susceptible individuals by \textbf{40\%}, \textbf{42\%}, \textbf{66\%}, \textbf{76\%}, \textbf{79\%}, and \textbf{88\%} with probability proportional to the sample size of the corresponding study \citepAP{puranik2021comparison, sheikh2021sars, lopez2021effectiveness, pouwels2021effect, fowlkes2021effectiveness}. See Table~\ref{tab:ve_params}.
        \item We have also applied uniform discrete distributions for both VE parameters (over three values for the VE in reducing viral load, and over six values for the VE in protecting against infections) and found the outcome to be of the same order of magnitude.
        \item Furthermore, our simulation for a single classroom requires specifying the vaccination status of the source case. This is simulated using a Bernoulli random variable with parameter $\mathbb{P}(\text{vaccinated} \mid \text{infected})$, which in turn can be deduced from $v_{\text{susceptible}}$ and $\beta_{\text{vaccinated}}$ using Bayes Rule. By definition,
        \begin{align*}
            \mathbb{P}(\text{infected}\mid \text{vaccinated}) = (1-v_{\text{susceptible}})\cdot \mathbb{P}(\text{infected}\mid \text{unvaccinated}).
        \end{align*}
        By Bayes Rule,
        \begin{align*}
            \mathbb{P}(\text{vaccinated} \mid \text{infected})
            &= \frac{\mathbb{P}(\text{infected}\mid \text{vaccinated})\cdot\mathbb{P}(\text{vaccinated})}{\mathbb{P}(\text{infected}\mid \text{vaccinated})\mathbb{P}(\text{vaccinated}) + \mathbb{P}(\text{infected}\mid \text{unvaccinated})\mathbb{P}(\text{unvaccinated})}\\
            &= \frac{(1-v_{\text{susceptible}})\cdot \beta_{\text{vaccinated}}}{1-v_{\text{susceptible}}\cdot \beta_{\text{vaccinated}}}.
        \end{align*}
        
    \end{enumerate}

\begin{table}[]
\caption{Vaccine efficacy estimates in the literature at the time of the analysis (summer 2021).}
\centering
\begin{tabular}{|l|l|l|l|}
\hline
 & Study & Mean & Sample size \\ \hline
\multirow{6}{*}{$v_{\text{susceptible}}$} & \cite{pouwels2021effect} & 40\% & 199,411 \\ \cline{2-4} 
 & \cite{puranik2021comparison} & 42\% & 22,064 \\ \cline{2-4} 
 & \cite{fowlkes2021effectiveness} & 66\% & 2,840 \\ \cline{2-4} 
 & \cite{puranik2021comparison} & 76\% & 21,179 \\ \cline{2-4} 
 & \cite{sheikh2021sars} & 79\% & 53,679 \\ \cline{2-4} 
 & \cite{andrews2021effectiveness} & 88\% & 15,871 \\ \hline
\multirow{3}{*}{$v_{\text{source}}$} & \cite{brown2021outbreak} & 0\% & 469 \\ \cline{2-4} 
 & \cite{harris2021effect} & 50\% & 96,898 \\ \cline{2-4} 
 & \cite{levine2021initial} & 71\% & 4,938 \\ \hline
\end{tabular}
\label{tab:ve_params}
\end{table}

    \item \textbf{Masking} ($\beta_{\text{masked}}$)
    
    Effective July 30, 2021, the University required all individuals, including fully vaccinated ones, to wear masks indoors \citepAP{cornell_indoor_mask_requirement}. We assume perfect compliance with this mandate. Thus we set $\beta_{\text{masked}}=\textbf{100\%}$.

\citetAP{jefferson2023physical} conducted a meta-analysis on the effectiveness of masking for reducing the spread of respiratory viruses during the COVID-19 pandemic. The results of the meta-analysis were inconclusive due to the limits of the primary-source evidence included in the review \citepAP{Soares-Weiser_2023}. The included studies were conducted in various settings (e.g., hospitals, communities, households), and masking compliance was low and the intensity and duration of interaction was high in many of these settings.  On the other hand, students are usually fully compliant with masking in the University's classrooms. Therefore, the review conducted by \citetAP{jefferson2023physical} would not have altered our decision to enforce masking.



    \item \textbf{Masking effectiveness} ($m$)
    
    We define masking effectiveness as the reduction in unmasked transmission probability due to masking. We assume two-way masking effectiveness, where both the source and susceptible are masked, to follow \textbf{Normal(0.855, 0.0536)}, from studies on one-way masking effectiveness in the literature.
    \begin{enumerate}
        \item We set 80\% as an optimistic estimate for one-way masking effectiveness.
        \begin{enumerate}
            \item Masking of infectious individuals: \cite{kumar2020utility} used fluid dynamics simulation and estimated that 12\% of the airflow carrying virus particles leaks around the side of a mask. \cite{wang2020reduction} observed that masking by the primary case and family contacts before the primary case developed symptoms was 79\% effective in reducing transmission.
            \item Masking of susceptible individuals: \cite{konda2020aerosol} experimentally measured the filtration efficiency of masks made from different materials and found that many materials could block particles larger than 0.3 micrometers with at least 96\% filtration efficiency. (\cite{morawska2009size} observed that most human expiratory activities generate droplets or aerosols with size larger than this.) \cite{doung2020case} reported that people wearing a mask all the time during contact with a COVID-19 patient are 84\% less at risk of infection. \cite{howard2021evidence} noted that wearing masks provides additional protection by preventing touching the nose and mouth, which is another vector of transmission.
        \end{enumerate}
        \item We set 50\% as a conservative estimate for one-way masking effectiveness. \cite{konda2020aerosol} observed that improper mask-wearing (e.g., having a gap between the face and the mask) can result in a large decrease in filtration efficiency. Unmasking temporarily to eat or drink would also reduce the protection. 
        \item Together, these imply that if masking is enforced, where the source and susceptible are both masked, transmission risk is reduced from the no-masking scenario by a multiplicative constant of $(1-80\%)^2 = 0.04$ to $(1-50\%)^2 = 0.25$.  Thus, for two-way masking, we set the prior on the risk reduction factor to be Normal(0.855, 0.0536), truncated to $[0,1]$. The untruncated distribution is designed such that [0.75, 0.96] is the 95\% symmetric confidence interval.
    \end{enumerate}

    \item \textbf{Prevalence in the University population} ($p$)
    
    We impose a prior distribution on prevalence, which is computed using the following procedure:
    \begin{itemize}
        \item We sample $N_{\text{infections}}$, the total number of infections over the entire semester from a distribution (details below).
        \item We conservatively assume each case is non-isolated and infectious for half a week, because surveillance testing requires that undergraduate students get tested twice a week on average. This is conservative since half a week is the maximum interval between tests, and the university took explicit measures to ensure consistent high compliance to testing \citepAP{cornell_test_noncompliance}.
        \item There are roughly 14 weeks in a semester. Thus, we approximate the prevalence at any time point during the semester as the total number of infected student-days divided by the total number of student-days in the semester:
        $$
        \frac{N_{\text{infections}} \cdot 3.5 \text{ days}} {n_{\text{UG}}\cdot (14\cdot 7) \text{ days}}.
        $$
        This is a simplifying assumption as we do not model the temporal change of prevalence as the semester proceeds.

        Before the start of the Fall 2021 semester, we assumed that $N_{\text{infections}}$ follows LogNormal(6.791, 0.413). This is derived by setting the mode to be 750 and the 97.5\% quantile to be 2000. Modeling results in early 2021 \citepAP{feb2021_report} show that a variant with more than twice the transmissibility of the original strain would lead to approximately 500 student infections, under the same masking and social distancing conditions as the Fall 2020 semester. To account for the relaxation of social distancing and masking (outside the classroom) in Fall 2021, we increased this number by 50\% and set 750 to be the mode. 

        By Sep. 19, 2021, we observed 488 student cases since the start of the semester and 32 student cases in the past week. Assuming the constant rate of 32 cases/week for the remaining 11 weeks of the semester, we would then expect to see 488 + 32*11 = 840 cases total. We repeated the simulations with $N_{\text{infections}}$ following the LogNormal(6.872, 0.372) distribution, derived by setting the mode to 840 and the 97.5\% quantile to be 2000, and got similar results. 
        At mid-October 2021, the University's public COVID-19 dashboard showed 22 weekly cases \citepAP{cornell_covid_dashboard}, so the assumed mode of 840 student cases was conservative.



        \item Prevalence is assumed to be proportional to $N_{\text{infections}}$ (which follows LogNormal(6.791, 0.413)), so it follows \textbf{LogNormal(-6.157, 0.413)}, where the mean is shifted from that of $N_{\text{infections}}$ by the log of the proportionality constant.
    \end{itemize}
    

    \item \textbf{Average hours an undergraduate spends in class in a semester} ($\tau_{\text{UG}}$)
    
    On average, students at the University are enrolled for 15 credits with 45 hours of course-related work per week, including both lecture, non-lecture time in classrooms (e.g., recitation), and time spent outside of class on homework and other coursework. We assume half of the 45 hours is spent in the classroom. Over a 14-week semester, a student spends $45 / 2 \times 14 = 315$ hours in the classroom. Thus we set $\tau_{\text{UG}}=\textbf{315}$.

    \item \textbf{Population sizes}
    \begin{enumerate}
        \item $n_{\text{UG}}=15,000$ \citepAP{university_facts}
        \item $n_{\text{faculty}} = 850$
        
        The Ithaca campus has roughly 1,700 faculty members \citepAP{university_facts}. We assume half of them teach undergraduate classes in a semester.
        
        \item $n_{\text{graduate}} = 3,120$ 
        
        In Fall 2020, there were 6,239 graduate students \citepAP{IRP_fall_2020}. We assume half of them work as TAs for classes or recitations. Note that some of these TAs may not interact face-to-face with students, e.g., if their primary responsibility is grading, and that the risk calculated is the average risk across all TAs, including these individuals. 
        \item $\beta_{\text{faculty}} = 2/3$
        
        On average, a course at the University has three meetings per week with two being lectures taught by faculty and one being a recitation led by a graduate student. Thus, we set this parameter to be 2/3.
    \end{enumerate}

    \item \textbf{Sufficient distancing of instructors} We assume that faculty and graduate student instructors are sufficiently distanced from the students during lecture that their risk of infection only arises from long-distance transmission. We assume social distancing is maintained during one-on-one discussions between instructors and students and that they do not add significantly to the total interaction time between instructors and students.
\end{enumerate}

\section{A3. Mathematical Model for the Risk of Transmission over Short and Long Ranges} \label{subsec:math_model}

In this section, we present the mathematical model for the transmission probability of COVID-19 depending on the relative location of the source case and susceptible individual. This is an important component in simulating infections in a single classroom, as described in Stage 1 in Section A1. 

Exposure to respiratory fluids is a major mode of transmission of SARS-CoV-2. An infectious source case releases the virus through exhalation of virus-containing respiratory fluids (e.g., through speaking, coughing, or sneezing). 
A susceptible person becomes exposed to the virus if they inhale the virus-containing aerosols or fine droplets, if the virus particles deposit on their mucous membranes via larger droplets, or if they touch a contaminated surface and then touch their mucous membranes \citepAP{cdc_transmission_modes}. We consider the first two possibilities here. (Hand sanitizers and disinfecting wipes are provided in all classrooms, which reduces the risk from the third mode of transmission.)
In particular, we model (1) \textit{long-range transmission} via aerosols and fine droplets that suspend in the air and (2) \textit{short-range transmission} via large droplets that eventually deposit after being emitted.
Under the assumption that instructors are at least 6 feet away from all students, instructors are only at the risk of long-range transmission, while students are subject to the risk of both. 
The modeling of both types of transmission relies on the exponential dose-response model, introduced below.

\subsection{Exponential Dose-response Model}

A dose-response model calculates the transmission probability as a function of \textit{dose}, the amount of virus particles a susceptible person is exposed to. In the exponential dose-response model \citepAP{watanabe2010development}, the transmission probability given dose $D$ takes the form
\begin{equation}
    \mathbb{P}(\text{transmission}) = 1 - \exp(-c\cdot D),\label{eq:exp_dose_response}
\end{equation}
where $c$ is a positive constant. Observe that $\mathbb{P}(\text{transmission})$ is concave in the dose $D$, a fact used in Section A1 (Simulation Tool) to observe that the increase in risk created by adding a second positive in the classroom is smaller than the increase in risk created by the first positive.

\subsection{Long-range Transmission}

We base our analysis of long-range transmission on the results from \cite{schijven2021quantitative}, who developed a model for predicting the transmission risk in an enclosed space due to aerosols only, under the assumption that emitted aerosols are dispersed across the entire room. The estimated risk depends on the aerosol-emitting activity (such as breathing, speaking, and singing), the level of ventilation of the room, the duration of interaction, and the virus concentration of the source case (measured in the number of virus particles per unit volume). One key property is that the risk due to aerosol transmission is uniform across all locations of the room, since the aerosols disperse quickly across space once emitted.

The risk of aerosol transmission over time $T$ is estimated by the exponential dose-response model:
\begin{align*}
    \mathbb{P}_{\text{aerosol}}(\text{transmission}, T) = 1 - \exp\left(-\frac{D(T)}{1440}\right),
\end{align*}
where $D$ denotes the \textit{dose}, i.e., the amount of virus particles that a susceptible person receives from the infectious person over time, and 1,440 is the estimated average number of virus copies to cause illness, according to \citetAP{schijven2021quantitative}. The dose depends on the number of virus particles emitted over time $T$, which we denote $N(T)$, as follows
\begin{align*}
    D(T) = N(T)\cdot \frac{\text{inhalation rate of the susceptible (volume / time)}}{\text{volume of the room}},
\end{align*}
where $N(T)$ depends on the type of aerosol-emitting activity, the viral load of the infectious person, the ventilation condition of the room, and the duration of interaction $T$. It is estimated that breathing emits aerosols containing 3,300 virus RNA copies per hour (assuming a nominal viral load $10^8$ copies per milliliter), while the value is higher for speaking and singing. We refer the readers to Equations (4) - (14) in \cite{schijven2021quantitative} for details of the calculation.

We implement a few additional calculations when deploying this model for the classroom simulation:
\begin{itemize}
    \item We average the risk over the distribution of the source viral load, which \cite{schijven2021quantitative} estimated to be log-normal. In particular, we compute the weighted average of transmission risk given that the source case viral load is $10^k$ copies per milliliter, for $k=5,6,\ldots,11$, with weights 0.12, 0.22, 0.3, 0.23, 0.103, 0.0236, 0.0034 respectively.  
    \item We take into account the effect of vaccination and masking, as described in the previous assumptions.
    \item We implement different ventilation conditions, namely no ventilation (a conservative estimate of the amount of ventilation in naturally-ventilated spaces), 1 air exchange per hour, and 3 air exchanges per hour. We model the amount of aerosols present in the room per hour as being reduced by a factor of two and four under the latter two conditions, respectively.
\end{itemize}


\subsection{Short-range Transmission}

\begin{table}[htbp]
\caption{Notation for functions and parameters in the model for transmission over short distances.}
\centering
\begin{tabular}{|l|l|}
\hline
\textbf{Notation} & \textbf{Meaning} \\ \hline
$\phi(r)$ & \begin{tabular}[c]{@{}l@{}}fraction of droplets that deposit at distance $\geq r$ meters\\ from their location of emission in 1D\end{tabular} \\ \hline
$\gamma_{1D}(r)$ & \begin{tabular}[c]{@{}l@{}}fraction of droplets deposited at distance $r$ away from the source\\ in 1D\end{tabular} \\ \hline
$\gamma_{2D}(r,\theta)$ & \begin{tabular}[c]{@{}l@{}}fraction of droplets deposited at $(r,\theta)$ away from the source\\ in the 2D model\end{tabular} \\ \hline
$\alpha$ &  parameter defining cone of exposure over $[-\alpha,\pi+\alpha]$\\ \hline
$\phi_{2D,ind}(r,\theta;\alpha)$ & \begin{tabular}[c]{@{}l@{}}fraction of droplets emitted by the source case that reach an \\ individual at $(r,\theta)$ assuming $\alpha$-cone of exposure\end{tabular} \\ \hline
$D(r,\theta,T)$ & \begin{tabular}[c]{@{}l@{}}dose of virus that a susceptible person at $(r,\theta)$ away \\ from the source receives throughout interaction duration $T$\end{tabular} \\ \hline
$N(x,y)$ & \begin{tabular}[c]{@{}l@{}}number of close contacts seated $x$ rows and $y$ columns \\ away from an index case\end{tabular} \\ \hline
$Y(x,y)$ & \begin{tabular}[c]{@{}l@{}}number of close contacts seated $x$ rows and $y$ columns \\ away from an index case that were later confirmed as positive\end{tabular} \\ \hline
$d(x,y)$ & \begin{tabular}[c]{@{}l@{}}distance between an index case and a close contact at \\ relative location $(x,y)$, computed from row-wise distance\\ $d_r(x,y)$ and column-wise separation $d_c(x,y)$\end{tabular} \\ \hline
$q_\alpha(x,y)$ & \begin{tabular}[c]{@{}l@{}}expected fraction of close contacts counted at $(x,y)$ \\ that are in the $\alpha$-cone of exposure\end{tabular} \\ \hline
\begin{tabular}[c]{@{}l@{}}$p_{c_2}((x,y),T\mid \text{in cone})$\end{tabular} & \begin{tabular}[c]{@{}l@{}}probability that a close contact in the cone of exposure \\ at $(x,y)$ is infected over duration $T$\end{tabular} \\ \hline
\end{tabular}
\label{tab:droplet_model_notation_table}
\end{table}

In this section, we first derive a mechanical model for the deposition of droplets over two-dimensional space over short distances. 
Based on the mechanical model, we derive an expression for the amount of droplets that a susceptible person at a certain location relative to the source receives (equivalently, the amount of droplets that reach the susceptible person spatially).
We then model the susceptible person's risk of infection due to exposure to viral droplets using the exponential dose-response model (Equation~\ref{eq:exp_dose_response}).
Finally, we estimate the model parameters from a dataset of transmissions on high speed trains in China \citepAP{hu2021risk}.
Table~\ref{tab:droplet_model_notation_table} summarizes the notation for the functions and parameters used in developing the model.


We first make a fundamental assumption for model tractability.
\begin{assumption}
\label{assu:uniform_virus_concentration}
The concentration of virus particles in droplets exhaled by a source case is uniform across all droplets of different sizes.
\end{assumption}


Assumption~\ref{assu:uniform_virus_concentration} allows us to use the \textit{volume} of viral droplets as a proxy for the amount of virus that a susceptible person is exposed to. This simplifies the calculation and allows us to better leverage existing results from the fluid dynamics literature.
We next assume that the transmission of droplets is not blocked by any obstacles.
\begin{assumption}
\label{assu:no_obstacle}
There are no obstacles between a source and a susceptible person, regardless of their locations. A susceptible person at distance $r$ from the source receives all droplets that would deposit at distance $r$ or further.
\end{assumption}


Using fluid dynamics modeling and experimental data, \cite{sun2020efficacy} estimated a function for the expected fraction of droplets that deposit at distance no less than $r$ meters from their location of emission, assuming all droplets travel in the same direction, over a distribution of droplet sizes from a typical cough: 
\begin{equation}
    \phi(r) = -0.1819\cdot \ln(r) + 0.43276.
\label{eq:frac_droplets_deposit}
\end{equation}
This formula is valid for $r\in [0.04, 10.8]$ meters, at the two ends of which $\phi(r)$ is equal to 1 and 0 respectively. We let $r_{\min}=0.04$ and $r_{\max}=10.8$.

If the source case exhaled all droplets in one direction, then a susceptible person at distance $r$ away in that exact direction would receive a fraction $\phi(r)$ of the droplets, while a susceptible person in all other directions would not receive any. 
In reality, however, the droplets emitted by an infectious individual may travel in multiple directions \citepAP{xie2009exhaled}, putting surrounding neighbors at different angles at risk. As the droplets are being spread out in space, the one-dimensional model in Equation~\ref{eq:frac_droplets_deposit} does not accurately capture the amount of droplets that reach one susceptible person in the vicinity. 

Thus, we extend this one-dimensional model to two dimensions. We design our 2D model of droplet deposition such that it depends on both the distance and angle of the susceptible person with respect to the source.
To ensure our 2D model is consistent with the 1D model, we make the following assumption: 

\begin{assumption}
\label{assu:1D_to_2D}
The fraction of droplets traveling beyond the entire radius-$r$ circle around the source is the same as $\phi(r)$, the fraction beyond distance $r$ in the 1D model.   
\end{assumption}

Now we derive the 2D model under Assumption~\ref{assu:1D_to_2D}. Let $\gamma_{1D}(r')$ denote the fraction of droplets deposited \textit{at} distance $r'$ away from the source in the 1D model. By definition, 
\begin{equation}
    \phi(r)=\int_{r}^{r_{\max}} \gamma_{1D}(r') dr'. \label{eq:1D_phi_gamma}
\end{equation}

Let $\gamma_{2D}(r',\theta)$ denote the fraction of droplets deposited at distance $r'$ and angle $\theta$ away from the source in the 2D model. By definition and Assumption~\ref{assu:1D_to_2D},
\begin{equation}
    \phi(r)= \int_{r}^{r_{\max}} \int_0^{2\pi} \gamma_{2D}(r',\theta) r' d\theta dr' = \int_{r}^{r_{\max}} \left[\int_0^{2\pi} \gamma_{2D}(r',\theta) d\theta \right] r' dr'.\label{eq:1D_2D_consistency_intermediate}
\end{equation}
The term in the bracket is the fraction of droplets that deposit over the entire circle of radius $r'$. From Equations~\ref{eq:1D_phi_gamma} and~\ref{eq:1D_2D_consistency_intermediate}, the 1D and 2D models should satisfy the following consistency condition:
\begin{equation}
    \int_0^{2\pi} \gamma_{2D}(r',\theta) d\theta = \frac{\gamma_{1D}(r')}{r'}.\label{eq:1D_2D_consistency}
\end{equation}

Our goal is to model the transmission risk that one susceptible person at distance $r$ and angle $\theta$ is subject to. Going from the fraction of droplets depositing over the entire circle, we next explicitly model the dependence of $\gamma_{2D}(r',\theta)$ on $\theta$. 

In realistic settings like classrooms or buses, a susceptible person could be seated in different directions from the source. We would naturally expect some directions to be riskier and others to be safer. For example, we would think of seats right behind the source as relatively safe, because the source is most likely facing and exhaling forward and the chair backs may block the droplets. On the other hand, it is possible that the source may turn their head from left to right when seated, so that droplets may potentially even reach someone sitting in rows behind them.

Based on these observations, we set up a ``cone of exposure" model that quantifies the dependence of risk on angle $\theta$. The cone of exposure, with parameter $\alpha$ (ranging from 0 to $\pi/2$), covers an arc of $(\pi/2+\alpha)$ degrees on both sides of the direction that the source case is facing. Hereafter, we call this an "$\alpha$-cone of exposure". An illustration is given in Figure~\ref{fig:cone_of_exposure}. The source case (purple) is facing up and emits droplets in directions ranging from angle $\alpha$ behind on their left to angle $\alpha$ behind on their right. Susceptible cases (blue) sitting within this cone are at the risk of receiving the droplets; susceptible cases sitting outside of this cone are not at risk.

\begin{figure}
    \centering
    \caption{Cone of exposure model. The source case is represented with a purple ``X" and the susceptible person is represented with a blue dot. The source emits virus uniformly over the cone extending from $-\alpha$ to $\pi + \alpha$. The susceptible person is located at distance $r$ and angle $\theta$ away from the source. They occupy an angle of $\Delta\theta$ that scales inversely with $r$.}
    \includegraphics[width=0.5\textwidth]{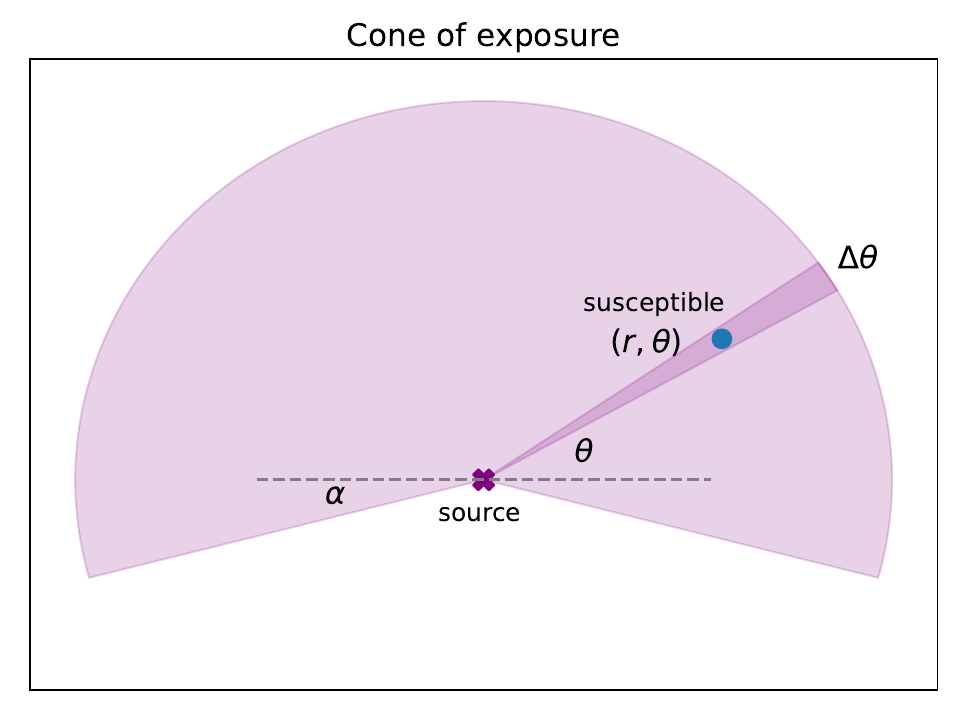}
    \label{fig:cone_of_exposure}
\end{figure}

We choose the right-hand direction of the source to be of angle 0 and measure $\theta$ counterclockwise. We let $\mathbbm{1}\{\theta \text{ in cone};\alpha\}$ be an indicator of whether $\theta$ is in the $\alpha$-cone of exposure, i.e., $\mathbbm{1}\{\theta \text{ in cone};\alpha\} = \mathbbm{1}\{\theta\in[-\alpha, \pi+\alpha]\}$. The amount of droplets depositing is positive for those in the cone and zero for those outside the cone. Next, we make a further assumption about droplet distribution within the cone.

\begin{assumption}
\label{assu:unif_dist_over_cone}
At the same distance, droplets are distributed uniformly over all angles in the cone of exposure.
\end{assumption}

This simplification aids the analysis, but in reality the distribution of droplets within the cone may depend on the angle in a more complicated way. We lack sufficient data to accurately estimate a complex angle-dependent droplet distribution model, and so a complex model might degrade accuracy rather than improve it. We thus adopt Assumption~\ref{assu:unif_dist_over_cone} in the spirit of regularization, understanding that the model we adopt approximates a more complex angle-dependent distribution model by replacing small droplet densities by 0 and large droplet densities by a constant.

With this cone-of-exposure model set up, we can model how the transmission risk depends on $\theta$.
We assume all individuals have comparable width. Then, for an individual at distance $r$ and angle $\theta$ from the source, they occupy an arc whose central angle $\Delta\theta(r)$ is approximately inversely proportional to $r$ and they receive all the droplets that would land in the sector across $[\theta, \theta+\Delta\theta(r)]$ and $[r, r_{\max}]$. 
Based on this insight, we derive $\phi_{2D,ind}(r,\theta;\alpha)$, the fraction of droplets emitted by the source case that reach such an individual. We call this quantity the \textit{droplet reception factor}.
\begin{align}
    \phi_{2D,ind}(r,\theta;\alpha) =~& \mathbbm{1}\{\theta \text{ in cone};\alpha\} \int_{r}^{r_{\max}} \int_{\theta}^{\theta+\Delta\theta(r)} \gamma_{2D}(r',\theta) r'd\theta dr'\nonumber\\
    =~& \mathbbm{1}\{\theta \text{ in cone};\alpha\} \int_{r}^{r_{\max}} \int_{0}^{2\pi} \gamma_{2D}(r',\theta) r'd\theta dr' \cdot \frac{\Delta\theta(r)}{\pi+2\alpha} \nonumber \\
    =~& \mathbbm{1}\{\theta \text{ in cone};\alpha\} \int_{r}^{r_{\max}} \gamma_{1D}(r') dr' \cdot \frac{\Delta\theta(r)}{\pi+2\alpha} \nonumber \\
    =~& \mathbbm{1}\{\theta \text{ in cone};\alpha\} \cdot \phi(r) \cdot \frac{\Delta\theta(r)}{\pi+2\alpha} \nonumber \\
    \propto ~& \mathbbm{1}\{\theta \text{ in cone};\alpha\} \cdot \phi(r)\cdot \frac{1}{r}, \label{eq:droplet_reception_rate_prop}
\end{align}
where the second equality follows from Assumption~\ref{assu:unif_dist_over_cone}, the third equality follows from Equation~\ref{eq:1D_2D_consistency}, and the fourth equality follows from Equation~\ref{eq:1D_phi_gamma}.
Later, we will show that having an undetermined proportionality constant does not affect our results as long as the dependence on $r$ and $\theta$ is modeled correctly.

\paragraph{Transmission probability calculation}
Given the expression for the droplet reception factor, we translate this to the probability of transmission using the exponential dose-response model in Equation~\ref{eq:exp_dose_response}.
The dose that a susceptible individual at $(r,\theta)$ from the source receives is proportional to the fraction of the source's virus particles that they receive. By Assumption~\ref{assu:uniform_virus_concentration}, this in turn is proportional to the droplet reception factor $\phi_{2D,ind}(r,\theta;\alpha)$.


Next, we observe that the dose is larger if the source case and susceptible person maintain the same relative location longer. We call this the \textit{duration of interaction}, denoted $T$. For example, $T$ is roughly one hour for a lecture. 
We make the following assumption about the droplet emission rate over time.
\begin{assumption}
\label{assu:constant_droplet_emission_rate}
The amount of droplets a source case emits per unit time is constant. 
\end{assumption}

Under Assumption~\ref{assu:constant_droplet_emission_rate}, the amount of droplets emitted, and hence the dose, is proportional to the duration of interaction $T$. Thus, the dose of virus particles that a susceptible person at $(r,\theta)$ away from the source receives can be expressed as
\begin{align*}
    D(r,\theta,T)=c_1\cdot \phi_{2D,ind}(r,\theta;\alpha) \cdot T,
\end{align*}
where $c_1$ captures the proportionality of the dose to the droplet reception factor and the duration of interaction.
We further absorb into $c_1$ the proportionality relation within the droplet reception factor (Equation~\ref{eq:droplet_reception_rate_prop}), yielding another constant $c_2$, and derive the final expression of the transmission probability for a susceptible person at $(r,\theta)$ away from a source case for a duration of interaction $T$:
\begin{equation}
    \mathbb{P}_{\text{droplet}}(\text{transmission},r,\theta,T;\alpha, c_2) = 1 - \exp\left(-c_2\cdot \mathbbm{1}\{\theta \text{ in cone};\alpha\} \cdot \frac{\phi(r)}{r} \cdot T\right).
    \label{eq:droplet_transmission_prob}
\end{equation}
As a sanity check, we can see that if a susceptible person at $(r,\theta)$ is not in the cone of exposure, the transmission probability is 0.


\paragraph{Parameter estimation}
The goal of this section is to first derive the likelihood for an empirically observed dataset based on the model above, and then find values of $\alpha$ and $c_2$ that maximize the likelihood. 

\cite{hu2021risk} studies 2,334 confirmed positive cases (``index cases") and 72,093 close contacts who had co-travel times of 0-8 hours from 12/19/2019 through 3/6/2020 on high-speed trains in China. They examine the association of attack rate with the spatial distance between pairs of index cases and close contacts.
Here, a ``close contact" was defined as a person who had co-traveled on a train within a 3-row seat distance of an index case within 14 days before symptom onset. 
We treat a close contact as equivalent to a susceptible individual in our model.
Table~\ref{tab:train_data_raw_counts} reports the number of close contacts, and, among them, those that were later confirmed as positive, at different seat locations with respect to an index case, within the period of study. We would like to derive the likelihood of observing the data in Table~\ref{tab:train_data_raw_counts}.

We now introduce additional notation for formalizing our likelihood model. 
Let $(x,y)$ denote the seat at $x$ rows and $y$ columns away from an index case, where $x$ ranges from 0 to 3 and $y$ ranges from 0 to 5. 
Let $N(x,y)$ denote the number of close contacts that are seated $x$ rows and $y$ columns away from an index case (hereafter we call this ``at relative location $(x,y)$" for abbreviation).
Let $Y(x,y)$ denote the number of close contacts at relative location $(x,y)$ from an index case that were later confirmed as positive.
Based on the train cabin layout give in Figure 1 in \cite{hu2021risk}, we calculate the separation between an index and a close contact at relative location $(x,y)$. In particular, let $d_r(x,y)$ and $d_c(x,y)$ denote the row-wise and column-wise distance in meters. For all $y$, $d_r(x,y)=0.9 x$; for all $x$, $d_c(x,y)$ is equal to 0, 0.5, 1.05, 1.6, 2.1, and 2.6 for $y=0,\ldots,5$ respectively. We then calculate $d(x,y)=\sqrt{d_r(x,y)^2 + d_c(x,y)^2}$ (Table~\ref{tab:train_data_distance_matrix}).


\begin{table}[]
\caption{Cases number and total number of passengers co-traveled with an index patient (\cite{hu2021risk}, Table S1). Entries are $Y(x,y)/N(x,y)$, where $x$ is the number of rows apart and $y$ is the number of columns apart.}\label{tab:train_data_raw_counts}
\centering
\begin{tabular}{|l|llllll|}
\hline
\multirow{2}{*}{{Rows apart}} & \multicolumn{6}{l|}{Columns apart} \\ \cline{2-7} 
 & \multicolumn{1}{l|}{0} & \multicolumn{1}{l|}{1} & \multicolumn{1}{l|}{2} & \multicolumn{1}{l|}{3} & \multicolumn{1}{l|}{4} & 5 \\ \hline
0 & \multicolumn{1}{l|}{--} & \multicolumn{1}{l|}{92/2605} & \multicolumn{1}{l|}{33/1996} & \multicolumn{1}{l|}{7/1845} & \multicolumn{1}{l|}{7/1825} & 3/1028 \\ \hline
1 & \multicolumn{1}{l|}{10/4791} & \multicolumn{1}{l|}{12/5084} & \multicolumn{1}{l|}{5/3664} & \multicolumn{1}{l|}{3/3464} & \multicolumn{1}{l|}{1/3525} & 1/1872 \\ \hline
2 & \multicolumn{1}{l|}{11/4386} & \multicolumn{1}{l|}{8/4751} & \multicolumn{1}{l|}{8/3429} & \multicolumn{1}{l|}{5/3212} & \multicolumn{1}{l|}{3/3250} & 3/1769 \\ \hline
3 & \multicolumn{1}{l|}{2/4026} & \multicolumn{1}{l|}{2/4395} & \multicolumn{1}{l|}{4/3110} & \multicolumn{1}{l|}{3/2945} & \multicolumn{1}{l|}{3/2970} & 1/1589 \\ \hline
\end{tabular}
\end{table}

\begin{table}[]
\caption{Distance $d(x,y)$ in meters, where $x$ is the number of rows apart and $y$ is the number of columns apart.}\label{tab:train_data_distance_matrix}
\centering
\begin{tabular}{|l|llllll|}
\hline
\multirow{2}{*}{Rows apart} & \multicolumn{6}{l|}{Columns apart} \\ \cline{2-7} 
 & \multicolumn{1}{l|}{0} & \multicolumn{1}{l|}{1} & \multicolumn{1}{l|}{2} & \multicolumn{1}{l|}{3} & \multicolumn{1}{l|}{4} & 5 \\ \hline
0 & \multicolumn{1}{l|}{--} & \multicolumn{1}{l|}{0.5} & \multicolumn{1}{l|}{1.05} & \multicolumn{1}{l|}{1.6} & \multicolumn{1}{l|}{2.1} & 2.6 \\ \hline
1 & \multicolumn{1}{l|}{0.9} & \multicolumn{1}{l|}{1.03} & \multicolumn{1}{l|}{1.38} & \multicolumn{1}{l|}{1.84} & \multicolumn{1}{l|}{2.28} & 2.75 \\ \hline
2 & \multicolumn{1}{l|}{1.8} & \multicolumn{1}{l|}{1.87} & \multicolumn{1}{l|}{2.08} & \multicolumn{1}{l|}{2.41} & \multicolumn{1}{l|}{2.77} & 3.16 \\ \hline
3 & \multicolumn{1}{l|}{2.7} & \multicolumn{1}{l|}{2.75} & \multicolumn{1}{l|}{2.90} & \multicolumn{1}{l|}{3.14} & \multicolumn{1}{l|}{3.42} & 3.75 \\ \hline
\end{tabular}
\end{table}

The data does not contain information about which \textit{direction} the close contacts were seated with respect to the index cases. 
However, directionality information is crucial for our modeling. 
Thus, we make the following assumption about the symmetry of distribution of close contacts.
\begin{assumption}
A close contact counted at $(x,y)$ is equally likely to have been seated in all possible directions at location $(x,y)$ away from the source.\label{assu:symmetric_seating_prob}
\end{assumption}

With slight abuse of notation, we let $(+x, +y)$ and $(+x, -y)$ denote the seats $x$ rows in front of the index case and $y$ columns to the right and left, respectively. Similarly, we let $(-x, \pm y)$ denote the seats $x$ rows behind the index case and $y$ columns to the right or left. 
Because we assume the cone of exposure has an angle larger than $\pi$, the seats $(+x, \pm y)$ are always in the cone of exposure. The seats $(-x, \pm y)$ are in the cone of exposure if and only if $\arctan\left({d_r(x,y)}/{d_c(x,y)}\right) \leq \alpha$.

Let $q(x,y;\alpha)$ denote the expected fraction of close contacts counted at $(x,y)$ (which could be at $(\pm x, \pm y)$) that are in the $\alpha$-cone of exposure. Based on Assumption~\ref{assu:symmetric_seating_prob},
\begin{align}
    q(x,y;\alpha) = \frac{1}{2} + \frac{1}{2} \mathbbm{1}\{\arctan\left({d_r(x,y)}/{d_c(x,y)}\right) \leq \alpha\}. \label{eq:q_frac_in_cone}
\end{align}
We can calculate $q(x,y;\alpha)$ for all possible $(x,y)$ pairs using Table~\ref{tab:train_data_distance_matrix}.



Next, let $p((x,y),T\mid \text{in cone};c_2)$ denote the probability that a close contact in the cone of exposure at location $(x,y)$ is infected. Based on Equation~\ref{eq:droplet_transmission_prob}, we model this as
\begin{align}
    p((x,y),T\mid \text{in cone};c_2) = 1 - \exp\left(-c_2 \cdot \frac{\phi(d(x,y))}{d(x,y)} \cdot \kappa_{\text{mask}} \cdot T\right), \label{eq:p_inf_prob_in_cone}
\end{align}
where we include an additional factor of masking effectiveness $\kappa_{\text{mask}}$, due to the fact that wearing a mask can reduce the amount of virus that a susceptible person is actually \textit{exposed to} (compared to the virus in the droplets that \textit{reach} where the person is), and that mask-wearing had been quite prevalent since late January of 2020 in China. \citeAP{konda2020aerosol} estimated that a poorly fitting mask made of cotton or silk will reduce virus dose by approximately 30\%. We heuristically select $\kappa_{\text{mask}}=0.8$, assuming that the dataset involved a mix of both masked and unmasked passengers\footnote{We constructed this model prior to conducting the simulations described in Appendix A1 and A2, so we used a point estimate for masking effectiveness. Nevertheless, the point estimate of 0.8 is well-aligned with the prior choice for masking effectiveness parameter $m$, discussed in Appendix A2.}. 
We keep this constant separate from $c_2$ so that $c_2$ solely captures the way transmission probability depends on the unreduced dose. 
We set $T$ to be 2.1 hours. This is the mean co-travel time over all pairs of index cases and close contacts in the data. Unfortunately, no information is given about the co-travel time for each individual pair.

We let $r(x,y;\alpha, c_2)$ denote the overall transmission probability at relative location $(x,y)$ under parameters $\alpha, c_2$. This is the product of the probability that a close contact at $(x,y)$ is in the cone of exposure (Equation~\ref{eq:q_frac_in_cone}) and the conditional probability that they become infected given that they are in the cone (Equation~\ref{eq:p_inf_prob_in_cone}):
\begin{align*}
    r(x,y;\alpha, c_2) = q(x,y;\alpha) \cdot p((x,y),T\mid \text{in cone};c_2).
\end{align*}
For the likelihood model, we assume the number of transmissions at each $(x,y)$ independently follows a binomial distribution:
\begin{align*}
    Y(x,y) \sim \text{Binomial}(N(x,y),r(x,y;\alpha, c_2)).
\end{align*}
Assuming the numbers of transmissions at each $(x,y)$ are independent, the likelihood of all observations $\mathcal{D} := \{N(x,y),Y(x,y)\}_{\substack{x=0,\ldots,3\\y=0,\ldots,5}}$ is given by:
\begin{align*}
    \mathcal{L}(\mathcal{D})=\prod_{(x,y)} \binom{N(x,y)}{Y(x,y)}\cdot r(x,y;\alpha, c_2)^{Y(x,y)}\cdot (1-r(x,y;\alpha, c_2))^{N(x,y)-Y(x,y)}.
\end{align*}
We compute the log-likelihood and let $c_3$ denote the constant term that does not depend on $\alpha$ or $c_2$:
\begin{align}
    \mathcal{\ell}(\mathcal{D})=\displaystyle\sum_{(x,y)} \left[Y(x,y)\log(r(x,y;\alpha, c_2)) + (N(x,y)-Y(x,y)) \log(1-r(x,y;\alpha, c_2))\right] + c_3.\label{eq:train_data_llh}
\end{align}
We next find values of $\alpha$ and $c_2$ that maximize the log-likelihood $\mathcal{\ell}(\mathcal{D})$.
Using a discretized grid search, we find that the log likelihood is maximized at $c_2 = 0.0135$ and multiple values of $\alpha$. We choose the largest possible value $\alpha=15$ degrees, or 0.26 radians, with the intention of being conservative. 

\subsection{Combining the Short and Long-range Transmission Risk}

In our simulation, the risk for a susceptible individual is 
\begin{align*}
    \max\left(\mathbb{P}_{\text{droplet}}(\text{transmission}), \mathbb{P}_{\text{aerosol}}(\text{transmission})\right),
\end{align*}
i.e., the larger of the predicted risk due to short and long-range transmission. This is justified by the fact that we inferred the parameters for the droplet model assuming all infections in the dataset result from short-range transmission; as such, the inferred model implicitly accounts for long-range transmission in the dataset. In practice, the short-range transmission risk only dominates the long-range transmission risk at short distances (approximately within three meters), while the risk due to aerosol is uniform across all locations.

Finally, we recall that the computed risk so far is based on studies on the original virus strain. Thus, we multiply the calculated risk by 2.4 to account for the increased transmissibility of the Delta variant, as described in Appendix A2. 

\bibliographystyleAP{informs2014} 
\bibliographyAP{references} 
\end{APPENDIX}

\end{document}